\documentclass[12pt]{article}
\pdfoutput=1

\usepackage{textcomp}
\usepackage{color}

\usepackage{putex}
\usepackage{autobreak}
\usepackage{graphicx}
\graphicspath{{plots/}}
\usepackage{tabularx}
\usepackage{caption}
\usepackage{amsmath}
\usepackage{bbm}
\usepackage{array}
\usepackage{subcaption}
\usepackage{epstopdf}
\usepackage{enumerate}
\usepackage{cite}
\usepackage{youngtab}
\usepackage{booktabs}
\usepackage{tensor}
\usepackage{slashed}
\usepackage[aligntableaux=center]{ytableau}
\usepackage[utf8]{inputenc}
\usepackage{rotating}
\usepackage{makecell}
\usepackage{multirow}
\usepackage{braket}
\usepackage{tikz}
\usepackage[
      colorlinks=true,
      linkcolor=blue,
      urlcolor=blue,
      filecolor=black,
      citecolor=red,
      linktocpage=true
      ]{hyperref}
\usepackage{bm}

\newcommand{\ed}{\,.}
\newcommand{\ec}{\,,}

\newcommand {\be} {\begin {equation}}
    \newcommand {\ee} {\end {equation}}
\newcommand {\nn} {\nonumber}
\newcommand {\bes} {\begin {equation*}}
    \newcommand {\ees} {\end {equation*}}

\newcommand{\es}[2] {\begin{equation} \label{#1} \begin{split} #2 \end{split} \end{equation}}

\newcommand{\cA}{{\mathcal A}}

\newcommand{\cC}{{\mathcal C}}

\newcommand{\cG}{{\mathcal G}}

\newcommand{\cN}{{\mathcal N}}
\newcommand{\cO}{{\mathcal O}}
\newcommand{\cP}{{\mathcal P}}
\newcommand{\cQ}{{\mathcal Q}}

\newcommand{\cT}{{\mathcal T}}

\newcommand{\cM}{{\mathcal M}}

\newcommand{\beq}{\begin{equation}}
	\newcommand{\eeq}{\end{equation}}

\def\ie{\begin{equation}\begin{aligned}}
  \def\fe{\end{aligned}\end{equation}}

\numberwithin{equation}{section}

\def\<{\langle}
\def\>{\rangle}

\def\AP#1{{\Orange [AP: #1]}}

\newcommand{\Dbar}{\bar{D}}

\makeatletter
\DeclareFontFamily{OMX}{MnSymbolE}{}
\DeclareSymbolFont{MnLargeSymbols}{OMX}{MnSymbolE}{m}{n}
\SetSymbolFont{MnLargeSymbols}{bold}{OMX}{MnSymbolE}{b}{n}
\DeclareFontShape{OMX}{MnSymbolE}{m}{n}{
	<-6>  MnSymbolE5
	<6-7>  MnSymbolE6
	<7-8>  MnSymbolE7
	<8-9>  MnSymbolE8
	<9-10> MnSymbolE9
	<10-12> MnSymbolE10
	<12->   MnSymbolE12
}{}
\DeclareFontShape{OMX}{MnSymbolE}{b}{n}{
	<-6>  MnSymbolE-Bold5
	<6-7>  MnSymbolE-Bold6
	<7-8>  MnSymbolE-Bold7
	<8-9>  MnSymbolE-Bold8
	<9-10> MnSymbolE-Bold9
	<10-12> MnSymbolE-Bold10
	<12->   MnSymbolE-Bold12
}{}

\let\llangle\@undefined
\let\rrangle\@undefined
\DeclareMathDelimiter{\llangle}{\mathopen}%
{MnLargeSymbols}{'164}{MnLargeSymbols}{'164}
\DeclareMathDelimiter{\rrangle}{\mathclose}%
{MnLargeSymbols}{'171}{MnLargeSymbols}{'171}

\makeatother
\allowdisplaybreaks

\let\bar\relax
\newcommand{\bar}[1]{\overline{#1}}

\newcommand{\dDisc}{\operatorname{dDisc}}
\newcommand{\cwa}{\scalebox{1}[-1]{${}_\curvearrowleft$}}
\newcommand{\cwar}{\scalebox{1}[-1]{${}_\curvearrowright$}}

\newcommand{\Bt}{B_{t}[\tau,J]}
\newcommand{\Bv}{B_{v}[\tau,J]}
\newcommand{\Btprot}{B^{\rm prot}_{t}}
\newcommand{\Ru}{\widetilde{R}_{u}[\tau,J]}
\newcommand{\Ruprot}{\widetilde{R}^{\rm prot}_{u}}
\newcommand{\Ruv}{\widetilde{R}_{v}[\tau,J]}
\newcommand{\Rt}{R_{t}[\tau,J]}
\newcommand{\Rtprot}{R^{\rm prot}_{t}}
\newcommand{\Rtv}{R_{v}[\tau,J]}
\newcommand{\Xtwo}{X^{\cN=2}_{s,t}[\tau,J]}
\newcommand{\Xfour}{X^{\cN=4}_{s,t}[\tau,J]}
\newcommand{\Itwo}{I[\tau, J]}
\newcommand{\Itwoprot}{I^{\rm prot}(\lambda)}
\newcommand{\Ifourtwo}{I_{2}[\tau, J]}
\newcommand{\Ifourtwoprot}{I^{\rm prot}_{2}(\lambda)}
\newcommand{\Ifourfour}{I_{4}[\tau, J]}
\newcommand{\Ifourfourprot}{I^{\rm prot}_{4}(\lambda)}
\newcommand{\Ifourp}{I_{p}[\tau, J]}
\newcommand{\Ifourpprot}{I^{\rm prot}_{p}(\lambda)}
\newcommand{\Psil}{\Psi_{\ell}[\tau,J]}
\newcommand{\Psilprot}{\Psi_{\ell}^{\rm prot}}
\newcommand{\Phil}{\Phi_{\ell}[\tau,J]}
\newcommand{\Philprot}{\Phi_{\ell}^{\rm prot}}

\let\originalleft\left
    \let\originalright\right
\renewcommand{\left}{\mathopen{}\mathclose\bgroup\originalleft}
    \renewcommand{\right}{\aftergroup\egroup\originalright}

\usepackage{xifthen}
\newcommand{\dif}[1][]{
    \ifthenelse{\NOT\isempty{#1}}{
        {\mathrm{d}}^{#1}
    }{
        \mathrm{d}
    }
}

\begin{document}

\preprint{}

\institution{Imp}{Abdus Salam Centre for Theoretical Physics, Imperial College London, London SW7 2AZ, UK}
\institution{Sissa}{SISSA, Via Bonomea 265, I-34136, Trieste, Italy}
\institution{Infn}{INFN, Sezione di Trieste, Via Valerio 2, I-34127, Trieste, Italy}

\title{Bootstrapping Holographic Theories\\ in the Planar Limit}

\authors{Shai M.~Chester,\worksat{\Imp} Daniele R. Pavarini,\worksat{\Imp}  and Alessandro Piazza \worksat{\Sissa,\Infn}}

\abstract{%
    We combine the numerical conformal bootstrap with supersymmetric localization to constrain the spectrum of holographic superconformal field theories in the planar large $N$ limit for finite 't Hooft coupling $\lambda$.
    We consider the stress tensor multiplet four-point function in $\mathcal{N}=4$ $SU(N)$ super-Yang-Mills (SYM), which is holographically dual to closed string scattering, as well as the flavor multiplet correlator in a certain \( 4d \) $\mathcal{N}=2$ $U\!Sp(2N)$ gauge theory with $SO(8)$ flavor symmetry, which is dual to open string scattering.
    In both cases, we combine dispersive sum rules with integrated constraints from supersymmetric localization to compute an upper bound on the scaling dimension of the lowest scalar single-trace operator for all $\lambda$.
    For $\mathcal{N}=4$ SYM, we find that our bounds are close to the known integrability result for all $\lambda$.
    For the $\mathcal{N}=2$ theory, where integrability is not yet available, we find that our bounds are close to weak and strong coupling predictions in the relevant regimes.
}
\date{}

\maketitle

\tableofcontents

\pagebreak

\section{Introduction and Summary}

Recent years have seen a revolution in our understanding of holographic conformal field theories (CFTs) through the interplay of three non-perturbative methods: integrability, supersymmetric localization, and the conformal bootstrap.
Progress was first made in the large $N$ planar limit for \( 4d \) $\mathcal{N}=4$ super-Yang-Mills (SYM) from integrability, which was able to efficiently compute the scaling dimensions of all single-trace operators for all values of the 't Hooft coupling $\lambda\equiv g_\text{YM}^2 N$ via the quantum spectral curve~\cite{Gromov:2013pga,Gromov:2023hzc}.
Similar results were also derived for the spectrum of 3d $\mathcal{N}=6$ ABJM theory~\cite{Bombardelli:2018bqz,Cavaglia:2014exa}, and most recently for the 2d D1-D5 CFT~\cite{Cavaglia:2021eqr,Ekhammar:2021pys,Ekhammar:2026ykk}.\footnote{In the 2d case, CFT data at strong coupling has also recently been computed using string field theory \cite{Cho:2018nfn}, and matched to integrability and the worldsheet bootstrap \cite{Alday:2026jkn}.}
However, integrability has not yet been able to compute OPE coefficients of the full spectrum,\footnote{%
    See~\cite{Basso:2015zoa} for some partial results on operators with very large charge under the R-symmetry.
} nor has it been able to compute scaling dimensions beyond the planar limit.

The first results beyond the planar limit and for OPE coefficients came from combining the numerical bootstrap~\cite{Rattazzi:2008pe,Beem:2013qxa} with integrated constraints from supersymmetric localization~\cite{Binder:2019jwn,Chester:2020dja}.
These methods were first applied to the stress tensor multiplet four-point function in $\mathcal{N}=4$ SYM at finite $N$ and complexified coupling $\tau\equiv \frac{4\pi}{g_\text{YM}^2}+i \frac{\theta}{2\pi}$~\cite{Chester:2021aun,Chester:2023ehi}, where upper bounds were computed on both the scaling dimension and OPE coefficient of the lowest dimension scalar long multiplet for finite $N$ and $\tau$.
The localization constraints in this case came from derivatives $\partial_m^4F\vert_{m=0}$ and $\partial_m^2\partial_\tau\partial_{\bar\tau}F\vert_{m=0}$ of the sphere free energy \( F \) deformed by the unique mass $m$,\footnote{%
    One can also consider derivatives of squashing, but these were shown to be equivalent to constraints from derivatives of $m$ and $\tau$~\cite{Chester:2025kvw}.
} which can be computed as a function of $\tau$ from localization~\cite{Pestun:2007rz,Alday:2023pet,Dorigoni:2021guq}.
These bounds matched weak coupling results~\cite{Velizhanin:2009gv,Eden:2012rr,Fleury:2019ydf,Eden:2016aqo,Goncalves:2016vir,Georgoudis:2017meq} at small $\lambda$,\footnote{%
    For scaling dimensions, the integrability results of~\cite{Gromov:2013pga} are consistent with weak coupling results.
} where the lowest dimension operator is single-trace, and strong coupling results at large $\lambda$~\cite{Chester:2019jas,Chester:2020vyz}, where the lowest dimension operator is double-trace.
Similar non-perturbative bounds from bootstrap and localization~\cite{Binder:2018yvd} were also computed for the stress tensor multiplet correlator in 3d $\mathcal{N}=8$ ABJM theory in~\cite{Chester:2024bij}.\footnote{%
    The numerical bootstrap has also been applied to $\mathcal{N}=6$ ABJM theory~\cite{Binder:2020ckj}, but integrated constraints from localization~\cite{Binder:2019mpb} were not yet imposed.
}
However, these bootstrap results could not tell us about single-trace operator CFT data at strong coupling, which is the novel contribution from integrability~\cite{Gromov:2013pga}, as single-trace operators are not the lowest dimension operators at strong coupling.\footnote{%
    The scaling dimensions of single-trace operators tend to scale as $\lambda^{1/4}$ at large $\lambda$, unlike double-trace operators, whose dimensions are integers at strong coupling.
}
As such, the strong coupling OPE coefficients of single-trace operators such as the Konishi remained out of reach.

These single-trace OPE coefficients were first computed in the planar limit by combining numerical bootstrap, localization, and integrability~\cite{Caron-Huot:2022sdy,Caron-Huot:2024loc}.\footnote{%
    See~\cite{Cavaglia:2021bnz,Cavaglia:2022qpg} for earlier work combining bootstrap and integrability to compute OPE coefficients of operators living on defects in $\mathcal{N}=4$ SYM.
}
In detail,~\cite{Caron-Huot:2022sdy} considered the stress tensor multiplet correlator of $\mathcal{N}=4$ SYM in the planar limit, where only single and double-trace operators appear in the conformal block expansion.
The OPE coefficients squared in this expansion are positive for single-trace operators, but not for double-trace operators, which is why the conventional conformal bootstrap cannot be applied.
Instead,~\cite{Caron-Huot:2022sdy,Caron-Huot:2024loc} applied a dispersion relation derived from Regge boundedness~\cite{Caron-Huot:2020adz,Penedones:2019tng,Caron-Huot:2021enk} to expand the correlator in Polyakov-Regge blocks, which in the planar limit only receive contributions from single-trace operators.
Thanks to the good Regge behavior of the theory at finite coupling, they derived anti-subtracted dispersive sum rules that further constrain the single-trace operators.
The authors then combined crossing symmetry applied to the Polyakov-Regge block expansion, anti-subtracted dispersive sum rules, the two integrated constraints from $\partial_m^4F\vert_{m=0}$ and $\partial_m^2\partial_\tau\partial_{\bar\tau}F\vert_{m=0}$, and the values of scaling dimensions of single-trace operators known from integrability~\cite{Gromov:2013pga}, in order to compute very tight upper and lower bounds on the OPE coefficients of these single-trace operators.
These bounds were saturated by both weak coupling results~\cite{Georgoudis:2017meq} and strong coupling results from the recently derived AdS Virasoro-Shapiro amplitude~\cite{Alday:2023mvu}.

Considering that the finite $N$ numerical bootstrap was able to compute bounds close to physical theories just by imposing localization, one might wonder whether the planar bootstrap can also achieve that without needing to input results from integrability.
This would be particularly relevant for holographic theories for which integrability is not yet available.
In this paper, we show that this is indeed possible.
We start by considering the planar stress tensor correlator in $\mathcal{N}=4$ SYM, using the same setup as~\cite{Caron-Huot:2022sdy,Caron-Huot:2024loc}.
We combine dispersion relations, crossing symmetry, and the two known localization constraints to compute upper bounds on the scaling dimension of the Konishi operator for all $\lambda$.
We find that our bounds are close to the known integrability result, both in the weak coupling regime as computed from Feynman diagrams~\cite{Velizhanin:2009gv} and in the strong coupling regime as computed from the AdS Virasoro-Shapiro amplitude~\cite{Alday:2023mvu}.

We then consider a holographic CFT for which finite $\lambda$ results from integrability are not yet known: a \( 4d \) $\mathcal{N}=2$ $U\!Sp(2N)$ gauge theory with four fundamental hypermultiplets, an antisymmetric hypermultiplet, and an $SO(8)\times SU(2)$ flavor symmetry.\footnote{%
    In the planar limit, this theory is identical (up to redefining $\lambda$ by a factor of two) to two other \( 4d \) $\mathcal{N}=2$ gauge theories: an $SU(N)$ gauge theory with two antisymmetric hypermultiplets and four fundamental hypermultiplets with flavor symmetry $U(4)\times SU(2)$, and a $U\!Sp(N)\times U\!Sp(N)$ gauge theory with one bi-fundamental hypermultiplet and four fundamental hypermultiplets with flavor symmetry $SO(4)\times SO(4)\times SU(2)$~\cite{Ennes:2000fu}. Our results thus also apply to these theories.
}
This theory is dual to type IIB string theory with four D7 branes, an O7 plane, and $N$ D3 branes~\cite{Aharony:1998xz}.
We consider the four-point function of $SO(8)$ flavor multiplets in this theory, which is dual to scattering of gluons on the D7 branes.
This correlator was previously considered at weak coupling in~\cite{Du:2024xbd}, at large $N$ and finite complexified coupling $\tau$ in~\cite{Behan:2023fqq, Chester:2025wti}, for $N=2$ and finite $\tau$ using the numerical bootstrap combined with localization in~\cite{Chester:2022sqb}, and in the planar 't Hooft limit at large $\lambda\equiv g_\text{YM}^2 N$ using the AdS Veneziano amplitude in~\cite{Alday:2024yax,Alday:2024ksp}.\footnote{%
    The IR value of $\lambda$ receives non-planar corrections relative to the UV value $ g_\text{YM}^2 N$~\cite{Behan:2023fqq}, but this is not relevant for the planar discussion in this work.
}
Here, we will study the correlator in the planar limit for finite $\lambda$ by combining the numerical bootstrap with the one integrated constraint $\partial_m^4F\vert_{m=0}$ that is non-trivial in this limit,\footnote{%
    In general there are three integrated constraints corresponding to the three quartic Casimirs of $SO(8)$~\cite{Chester:2022sqb}, but only one gives a non-trivial constraint in the planar 't Hooft limit.
} which was computed using localization at large $N$ and finite $\lambda$ in~\cite{Behan:2023fqq}.
As in the planar $\mathcal{N}=4$ case, we eliminate double-trace operators by deriving dispersion relations and expanding the correlator in the resulting Polyakov-Regge blocks.
We also include anti-subtracted dispersive sum rules that come from the soft Regge behavior of the dual string scattering amplitude.
We then compute bounds on the scaling dimension of the lowest scalar long multiplet for finite $\lambda$, which are close to weak coupling results~\cite{Du:2024xbd} as well as strong coupling results from the AdS Veneziano amplitude~\cite{Alday:2024yax,Alday:2024ksp} in the relevant regimes.

The rest of this paper is organized as follows.
In Section~\ref{setup} we discuss the correlators we consider for the $\mathcal{N}=4$ and $\mathcal{N}=2$ theories, including their block expansion, integrated constraints, and dispersion relations.
For $\mathcal{N}=4$, this is a review of~\cite{Caron-Huot:2022sdy,Caron-Huot:2024loc}, while for $\mathcal{N}=2$ the dispersion relations are new.
In Section~\ref{numBoot}, we then show how to use these results to numerically bootstrap the theories, and present the resulting bounds on the scaling dimensions of scalar operators for general $\lambda$.
We conclude in Section~\ref{conc} with a review of our results and a discussion of future directions.
Technical details of the calculations are given in the various Appendices.

\section{Correlators}
\label{setup}

We study two four-point functions that we discuss in parallel: the correlator of stress tensor superprimaries in $\mathcal{N}=4$ SYM, which is holographically dual to closed string scattering, and the correlator of $SO(8)$ flavor multiplet superprimaries in the $\mathcal{N}=2$ theory with $SO(8)\times SU(2)$ flavor symmetry, which is dual to open string scattering.
We start by reviewing the non-perturbative block expansion of each correlator, and specialize to the planar limit.
We review analytic results at weak and strong coupling for each correlator.
We then discuss dispersion relations and the Polyakov-Regge block expansion for each correlator, which for $\mathcal{N}=2$ is a new derivation.
Finally, we discuss how to impose constraints on the planar correlator from crossing symmetry, dispersion relations, and localization.
In general, we give more details for the $\mathcal{N}=2$ case, as all the ingredients for $\mathcal{N}=4$ SYM are already present in the literature~\cite{Caron-Huot:2022sdy,Caron-Huot:2024loc}.

\subsection{Conformal block expansions}
\label{correlators}

We will first discuss the four-dimensional $\mathcal{N}=2$ $U\!Sp(2N)$ gauge theory with flavor group $G_F=SO(8)\times SU(2)$. The matter content of the theory consists of eight half-hypermultiplets in the fundamental representation of the gauge group, transforming in the fundamental of $SO(8)$; and two half-hypermultiplets in the antisymmetric representation of the gauge group, transforming in the fundamental of $SU(2)$. Consider the superprimary $\cO^{a;\alpha_1\alpha_2}(x)$ of the \( SO(8) \) flavor multiplet, which is a $\Delta=2$ scalar with indices \nobreak{$a,b,\ldots \in\{1,\dots,28\}$} for the \( \mathbf{28} \) adjoint of $SO(8)$ and $\alpha,\beta,\ldots\in\{1,2\}$ for the fundamental of the $SU(2)_R$ R-symmetry subgroup.\footnote{%
    Our setup is insensitive to the $SU(2)$ flavor symmetry and to the $U(1)$ part of the R-symmetry.
}
It is convenient to write the R-symmetry dependence of this so-called \emph{moment map operator} using polarization vectors as
\es{momentmap}{
    \cO^a(x;y)=\cO^{a;\alpha_1\alpha_2}(x)\,y^{\beta_1}y^{\beta_2}\epsilon_{\alpha_1\beta_1}\epsilon_{\alpha_2\beta_2}\,.
}
We study the four-point function
\es{Gdef}{
    \langle \cO^a(x_1;y_1)\cO^b(x_2;y_2)\cO^c(x_3;y_3)\cO^d(x_4;y_4)\rangle=\frac{(y_1\!\cdot y_3)^2(y_2\!\cdot y_4)^2}{x_{13}^4 x_{24}^4}\,G^{abcd}(U,V;\alpha)\ec
}
where we define the position-space cross ratios
\es{}{U=z\bar z=\frac{x_{12}^2x_{34}^2}{x_{13}^2x_{24}^2}\,,\qquad V=(1-z)(1-\bar z)=\frac{x_{14}^2x_{23}^2}{x_{13}^2x_{24}^2}\,,
}
and R-symmetry cross ratio
\es{}{\alpha=\frac{(y_1\cdot y_2)(y_3\cdot y_4)}{(y_1\cdot y_3)(y_2\cdot y_4)}\,.}
The superconformal Ward identity~\cite{Dolan:2001tt,Nirschl:2004pa} can be solved by splitting the correlator into a free part and an interacting part that is an R-symmetry singlet:
\es{Gsplit}{
    G^{abcd}(U,V;\alpha)=G^{abcd}_{\rm free}(U,V;\alpha)+(\alpha-z)(\alpha-\bar z)\,T^{abcd}(U,V)\ec
}
where
\es{Gfree}{
    G^{abcd}_{\rm free}&=\frac{\alpha^2}{U^2}\delta^{ab}\delta^{cd}+\delta^{ac}\delta^{bd}+\frac{(\alpha-1)^2}{V^2}\delta^{ad}\delta^{bc}+\frac{2\alpha}{kU}\tr(T^aT^bT^dT^c)\\
    &\quad +\frac{2\alpha(\alpha-1)}{kUV}\tr(T^aT^bT^cT^d)+\frac{2(1-\alpha)}{kV}\tr(T^aT^dT^bT^c)\ed
}
Here, $k=4N$ is the flavor central charge~\cite{Aharony:2007dj}, and the interacting part $T^{abcd}(U,V)$ carries all the dynamical information.
We can expand the interacting part in superblocks as
\es{block}{
    T^{abcd}(U,V)=T^{abcd}_\text{short}(U,V)+\sum_{r\in{\bf 28}\otimes {\bf 28}}P_r^{abcd}\sum_{J,\tau\geq2}\lambda^2_{\tau,J,r}U^{-3}g_{\tau+2,J}(U,V) \ed
}
The first term in~\eqref{block} contains all the protected multiplets whose CFT data does not depend on the 't Hooft coupling $\lambda$, and takes the explicit form~\cite{Chester:2022sqb}
\begin{equation}
    \label{Tprot}
    T^{abcd}_\text{short}=\frac{1}{z^2\bar z^2}\frac{1}{z-\bar z}\left[\log(1-z)\left( -2\delta^{ab}\delta^{cd}\frac{\bar z+\log(1-\bar z)}{c\,\bar z}+\bar z^2\delta^{ac}\delta^{bd}+\frac{\bar z^2\delta^{ad}\delta^{bc}}{(1-\bar z)^2} \right.\right.\phantom{aaaaa}
\end{equation}
\vspace{-0.5cm}
$$ \phantom{aa}\left.\left. +\frac{2\bar z f^{ade}f^{bce}}{k(\bar z-1)}+\frac{2\bar z f^{ace}f^{bde}}{k}\right) -(z\leftrightarrow\bar z)\right]-\frac{1}{UV^2}\delta^{ad}\delta^{bc}-\frac{1}{U}\delta^{ac}\delta^{bd}-\frac{2}{kUV}\tr(T^aT^dT^bT^c)\,,$$
with $c=\frac{N^2}{2}+\frac{3N}{4}-\frac{1}{12}$ the central charge of the theory~\cite{Aharony:2007dj}.

The sum in~\eqref{block} is over long multiplets of spin \( J \) and twist \( \tau \) that are R-symmetry singlets and transform in each irrep \( r \) appearing in the tensor product of two adjoints of $SO(8)$
\es{repdecomp}{
    \mathbf{28}\otimes\mathbf{28}=\mathbf{1}\oplus\mathbf{28}\oplus\mathbf{35}_c\oplus\mathbf{35}_s\oplus\mathbf{35}_v\oplus\mathbf{300}\oplus\mathbf{350}\,.
}
The spin of the long multiplet is even for irreps ${\bf1}, {\bf 35}_s, {\bf 35}_c,  {\bf 35}_v, {\bf 300}$ in the symmetric product, and odd for irreps in the antisymmetric product ${\bf28}, {\bf 350}$. Crossing symmetry gives seven equations that relate CFT data in the seven irreps, and is given explicitly in~\cite{Chester:2022sqb}.
We will not use these seven crossing equations in this work, as we consider the planar limit, where we will find that crossing simplifies drastically.

We will also find it convenient to consider the Mellin transform of the correlator, which we define as~\cite{Mack:2009mi,Penedones:2010ue}
\es{MellinN2}{
    T^{abcd}(U,V)= \int\frac{\dif{s}\,\dif{t}}{(4\pi i)^{2}}\,U^{\frac s2-2}V^{\frac t2-2}\,\Gamma\big(2-\tfrac s2\big)^2\,\Gamma\big(2-\tfrac t2\big)^2\,\Gamma\big(2-\tfrac u2\big)^2\,M^{abcd}(s,t)\,,
}
where $u=6-s-t$.

We next discuss the correlator of the stress tensor multiplet superprimary in $\mathcal{N}=4$ $SU(N)$ SYM, following e.g.~\cite{Caron-Huot:2024loc}.
The superprimary $\phi(x;\tilde y)$ is a $\Delta=2$ scalar operator that transforms in the rank-2 traceless symmetric irrep of the R-symmetry $SO(6)_R$, where $\tilde y$ here is a polarization vector for $SO(6)_R$. We study the four-point function
\es{Gdef4}{
    \langle \phi(x_1;\tilde y_1)\phi(x_2;\tilde y_2)\phi(x_3;\tilde y_3)\phi(x_4;\tilde y_4)\rangle=\frac{\tilde y_{13}^4\tilde y_{24}^4}{x_{13}^4 x_{24}^4}\,\cG(U,V;\alpha,\bar\alpha)\ec
}
where we define R-symmetry cross ratio
\es{}{\alpha\bar\alpha=\frac{\tilde y_{12}^2\tilde y_{34}^2}{\tilde y_{13}^2\tilde y_{24}^2}\,,\qquad (1-\alpha)(1-\bar\alpha)=\frac{\tilde y_{23}^2\tilde y_{14}^2}{\tilde y_{13}^2\tilde y_{24}^2}\,.}
The superconformal Ward identity~\cite{Dolan:2001tt} can be solved by splitting the correlator into a free part and an interacting part \( \cT(U,V) \) that is an R-symmetry singlet:
\es{Gsplit4}{
    \cG(U,V;\alpha,\bar\alpha)=\mathcal{G}_{\rm free}(U,V;\alpha,\bar\alpha)+\frac{1}{c}(z-\alpha)(z-\bar\alpha)(\bar z-\alpha)(\bar z-\bar\alpha)\,\cT(U,V)\,,
}
where $c=\tfrac{N^2-1}{4}$ is the stress tensor central charge (and conformal anomaly), and
\es{Gfree4}{
    \cG_{\rm free}=1
    +\frac{(\alpha\bar{\alpha})^2}{U^2}
    +\frac{\bigl[(1-\alpha)(1-\bar{\alpha})\bigr]^2}{V^2}
    +\frac{1}{c}
    \left[
        \frac{\alpha\bar{\alpha}}{U}
        +\frac{(1-\alpha)(1-\bar{\alpha})}{V}
        +\frac{\alpha\bar{\alpha}(1-\alpha)(1-\bar{\alpha})}{UV}
    \right].
}
We can expand the interacting part in superblocks as
\es{block4}{
    \cT(U,V)=\cT_\text{short}(U,V)+\sum_{J,\tau\geq2}\lambda^2_{\tau,J}U^{-4}g_{\tau+4,J}(U,V)\,,
}
where the first term contains all the protected multiplets whose CFT data does not depend on $\lambda$ (and whose explicit form can be found e.g.\ in~\cite{Beem:2016wfs}), while the sum is over all R-symmetry singlet long multiplets of twist \( \tau \) and \emph{even} spin $J$.
Crossing symmetry constrains $\cT(U,V)$ as
\es{cross4}{
    \cT(U,V)=\cT(V,U)=U^{-4}\cT(1/U,V/U)\,.
}
The Mellin amplitude for the stress tensor correlator can be defined as
\es{MellinN4}{
    \cT(U,V)=\int\frac{\dif{s}\,\dif{t}}{(4\pi i)^{2}}\,U^{\frac s2-4}V^{\frac t2-4}\,\Gamma\big(4-\tfrac s2\big)^2 \, \Gamma\big(4-\tfrac t2\big)^2 \, \Gamma\big(4-\tfrac{\tilde u}{2}\big)^2 \,\cM(s,t)\,,
}
where $\tilde u=16-s-t$.
Crossing symmetry in Mellin space then acts by simply permuting $s,t,\tilde u$.

\subsection{Planar limit}
We now specialize to the planar limit and explore the consequences for the correlator in the $\mathcal{N}=2$ case. The \( \cN = 4 \) case is more trivial and was reviewed in~\cite{Caron-Huot:2022sdy}.\footnote{%
    In our notation \( {\cT(U,V)\lvert_{\text{here}}} = {\mathcal{H}(u,v)\lvert_{\text{there}}} \).
}

In the planar limit, the color structure of the interacting part of the $\mathcal{N}=2$ moment map correlator is fixed by requiring that it reproduces the color structure of a tree-level gluon amplitude in the $\mathrm{AdS}$ holographic theory~\cite{Alday:2021odx}:
\begin{equation}\label{colorstrip}
      T^{abcd}(U,V)= \tr(T^aT^bT^cT^d)\,T(U,V) \!+\! \tr(T^aT^dT^bT^c) \tfrac{1}{U^{3}}T\big(\tfrac{V}{U},\tfrac1U\big)\! +\! \tr(T^aT^cT^dT^b)\tfrac{1}{V^{3}}T\big(\tfrac1V,\tfrac UV\big) \ed
\end{equation}

We are thus left with a single \emph{color-stripped} correlator $T(U,V)$ that captures all the dynamics.\footnote{Note that this does not apply away from the planar limit, since at sub-leading orders in \( N \) the correlator carries other flavor structures corresponding to multi-trace operators.}
Crossing symmetry acts by permuting the three color orderings in~\eqref{colorstrip} among themselves, so that all of its content is encoded in the single condition
\es{}{T(U,V)=T(V,U)\,.}
The color-stripped correlator has all the analytic properties of the correlator $T^{abcd}$; in fact, it is just the linear combination
\es{lincomb}{
    T(U,V)=\frac{1}{2}\left(T^{1122}(U,V)+T^{1221}(U,V)-T^{1212}(U,V)\right)\ed
}
The color structure of the Mellin amplitude can be analogously simplified to
\begin{equation}\label{colorstripMellin}
    M^{abcd}(s,t) = \tr(T^aT^bT^cT^d)M(s,t)+\tr(T^aT^dT^bT^c)M(t,u)+\tr(T^aT^cT^dT^b)M(u,s) \ec
\end{equation}
where crossing now implies that
\begin{equation}
    M(s,t) = M(t,s) \ed
\end{equation}
This decomposition is graphically represented in Figure~\ref{fig:colorordering}.

\begin{figure}[t]
    \centering
    \begin{tikzpicture}[line width=0.9pt]
        \foreach \dx in {0,4,8}{
            \begin{scope}[shift={(\dx,0)}]
                \fill[gray!15] (0,0) circle (0.7);
                \foreach \ang in {135,45,-45,-135}{
                    \fill[gray!15] ({\ang+3}:0.7) -- ++(\ang:0.15)
                    -- ++({\ang-90}:{2*0.7*sin(3)}) -- ++({\ang+180}:0.15) -- cycle;
                }
            \end{scope}
        }
        \begin{scope}[shift={(0,0)}]
            \foreach \ang/\lab in {135/a, 45/d, -45/c, -135/b}{
                \draw ({\ang+3}:0.7) -- ++(\ang:0.15);
                \draw ({\ang-3}:0.7) -- ++(\ang:0.15);
                \node at (\ang:1.1) {$\lab$};
            }
            \foreach \a/\b in {135/45, 45/-45, -45/-135, -135/-225}{
                \draw ({\a-2.5}:0.7) arc ({\a-2.5}:{\b+2.5}:0.7);
            }
        \end{scope}
        \begin{scope}[shift={(4,0)}]
            \foreach \ang/\lab in {135/a, 45/c, -45/b, -135/d}{
                \draw ({\ang+3}:0.7) -- ++(\ang:0.15);
                \draw ({\ang-3}:0.7) -- ++(\ang:0.15);
                \node at (\ang:1.1) {$\lab$};
            }
            \foreach \a/\b in {135/45, 45/-45, -45/-135, -135/-225}{
                \draw ({\a-2.5}:0.7) arc ({\a-2.5}:{\b+2.5}:0.7);
            }
        \end{scope}
        \begin{scope}[shift={(8,0)}]
            \foreach \ang/\lab in {135/a, 45/b, -45/d, -135/c}{
                \draw ({\ang+3}:0.7) -- ++(\ang:0.15);
                \draw ({\ang-3}:0.7) -- ++(\ang:0.15);
                \node at (\ang:1.1) {$\lab$};
            }
            \foreach \a/\b in {135/45, 45/-45, -45/-135, -135/-225}{
                \draw ({\a-2.5}:0.7) arc ({\a-2.5}:{\b+2.5}:0.7);
            }
        \end{scope}
        \node at (-2.2,0) {$\cA\;=$};
        \node at (2,0) {$+$};
        \node at (6,0) {$+$};
    \end{tikzpicture}
    \caption{Color decomposition of the four-point amplitude of moment map operators in the planar limit. The amplitude is a sum over the three inequivalent cyclic orderings $(abcd)$, $(adbc)$ and $(acdb)$ of the flavor indices along the boundary of the disk, each multiplying the same color-stripped amplitude evaluated at the corresponding cross ratios. The solid lines indicate the flavor flow.}
    \label{fig:colorordering}
\end{figure}

In the planar limit the OPE of the reduced correlator receives contributions only from single- and double-trace operators.
We are interested in studying the single-trace spectrum as the OPE data in this sector depends non-trivially on the 't Hooft coupling at the leading order in $N$, and single-trace operators are dual to heavy string states.
Single-trace operators contribute to $T$ only in the singlet $\mathbf{1}$, the adjoint $\mathbf{28}$, and the symmetric traceless $\mathbf{35}_v$ representations appearing in the decomposition~\eqref{repdecomp}.
Requiring that the OPE reproduces the color structure~\eqref{colorstrip} and imposing crossing symmetry, we find that the operators in the singlet and $\mathbf{35}_v$ representations are related.
For each single-trace operator in the $\mathbf{35}_v$, there is a singlet with the same dimension and spin, and the OPE coefficients are related by
\es{ratio7over3}{
    \lambda^2_{\mathbf{1},\tau,J}=\frac{7}{3}\lambda^2_{\mathbf{35}_v,\tau,J}\,.
}
Therefore, the single-trace contributions to the OPE can be organized into two terms: operators in the \( \mathbf{35}_{v} \) representation with even spin \( J \), and operators in the \( \mathbf{28} \) representation with odd spin \( J \).
We derive the OPE for the color-stripped correlator by applying~\eqref{lincomb} to~\eqref{block}. Using~\eqref{ratio7over3} and the expressions for the projectors in~\eqref{projbasis}, we find
\es{colorstrippedOPE}{
T(U,V)=T_\text{short}(U,V)+\frac{1}{12 k}\sum_{\text{single-trace}} &\lambda^2_{\mathbf{35}_v,\tau,J}U^{-3}g_{\tau+2,J}(U,V)+\lambda^2_{\mathbf{28},\tau,J}U^{-3}g_{\tau+2,J}(U,V)\\
+&\frac{1}{2}\sum_{\text{double-trace}}\lambda^2_{r,\tau,J}U^{-3}g_{\tau+2,J}(U,V)\ec
}
where
\es{Tshort}{T_\text{short}(U,V)=\frac{1}{z-\bar{z}}\left[\frac{\log(1-z)}{z^2(1-\bar{z})}\left(\frac{(2-\bar{z})\bar{z}}{2(1-\bar{z})}-\frac{1}{k}\frac{2}{\bar{z}}\right)-(z\leftrightarrow \bar{z})\right]+\frac{1}{2U}-\frac{1}{2UV^2}\,,
}
and we extracted a factor of $k$ from the OPE coefficients of the single-trace operators so that they are finite in the planar limit.

\subsection{Weak and strong coupling}
\label{weakStrong}
The correlators we consider in the planar limit have been studied at both weak and strong coupling.
For the $\mathcal{N}=2$ correlator, the weak coupling result is known to 1-loop order and takes the form of a box diagram~\cite{Du:2024xbd}\footnote{%
    We fix a typo in the overall normalization of the 1-loop correlator in~\cite{Du:2024xbd}. This normalization is checked by comparing to the localization results via the integrated constraint in Appendix~\ref{weak}.
}
\es{weakN2}{
    M(s,t)\Big|_{\text{1-loop}}=-\frac{2\lambda}{\pi^2 k}\frac{1}{(s-2)^2(t-2)^2}\ec\qquad T(U,V)\Big|_{\text{1-loop}}=-\frac{\lambda}{8\pi^2 k}\frac{\Dbar_{1,1,1,1}(U,V)}{UV}\,.
}
From this, we can read off all CFT data that appears in the four-point function; in particular, for the leading Regge trajectory we find
\es{}{
      \tau_r=2+\lambda \gamma^{\mathcal{N}=2}_J\qquad\lambda^2_{r,\tau,J}=C^{(0),\,\mathcal{N}=2}_J+\lambda C^{(1),\,\mathcal{N}=2}_J\,,
}
\es{}{\gamma^{\mathcal{N}=2}_J=\frac{H_{J+1}}{4 \pi ^2} \qquad
C^{(0),\,\mathcal{N}=2}_J=\frac{3\times 4^{1-J} (2)_J}{(3/2)_J}}
\es{}{C^{(1),\,\mathcal{N}=2}_J=-\frac{3 \Gamma (J+2)^2 \Big[(H_J)^2-3 H_{J+1} H_J+2 H_{J+1} H_{2 J+1}+H_J^{(2)}\Big]}{\pi ^2 \Gamma (2 J+3)}}
where $H_J^{(2)}$ is the harmonic number of order 2.
From here on we will only be interested in the lowest dimension single-trace scalar operator in the singlet irrep, i.e.\ the Konishi operator, whose scaling dimension at weak coupling is
\es{weakData}{
    \Delta_{K}^{\mathcal{N}=2}=2+\frac{\lambda}{4\pi^2}+O(\lambda^2)\,.
}
The leading term at strong coupling (i.e.\ $\lambda\to\infty$) is the tree-level gluon exchange, which is fixed by the analytic bootstrap~\cite{Alday:2021odx}. From the result in position space
\es{Tgluonfull}{T_{\text{gluon}}^{abcd}(U,V)=-\tfrac{2}{k}(U^{-1}\Dbar_{1,2,2,3}(U,V)(\mathtt{t}_4-\mathtt{t}_5)-V^{-1}\Dbar_{3,1,2,2}(U,V)\mathtt{t}_5+U^{-1}V^{-1}\mathtt{t}_6)\,,}
where the tensor structures $\mathtt{t}_i$ are defined in Appendix~\ref{app:mack}, we find, using~\eqref{lincomb},
\es{strongN2}{
    M_{\rm gluon}(s,t)=-\frac{8}{k(s-2)(t-2)}\ec\qquad T_{\rm gluon}(U,V)=-\frac{2}{k UV}\Dbar_{2,1,2,3}(U,V)\ed
}
The scaling dimensions of single-trace operators go as $\lambda^{1/4}$ at large $\lambda$ in the planar limit, so they cannot be read off from a conventional conformal block expansion of this correlator. Instead, they can be extracted from the AdS Veneziano amplitude, which was used in~\cite{Alday:2024yax,Alday:2024ksp} to compute the first two corrections, giving
\es{strongData}{
    \Delta_K^{\mathcal{N}=2}=\lambda^{1/4}+\frac12\lambda^{-1/4}+\left(\frac18-\frac72\zeta_3\right)\lambda^{-3/4}+O(\lambda^{-5/4})\,.
}

The $\mathcal{N}=4$ correlator is much more studied, and is in fact known to 4-loop order~\cite{Eden:2011we,Eden:2012tu}.
We will only make use of the 1-loop term, which takes the form
\es{MweakN4}{
    \cM(s,t)\Big|_\text{1-loop}=\frac{-\lambda}{8\pi^2\big(\tfrac s2-3\big)^2\big(\tfrac t2-3\big)^2\big(\tfrac{\tilde u}{2}-3\big)^2}\,,\qquad \cT(U,V)\Big|_{\text{1-loop}}=-\frac{\lambda}{8\pi^2 }\frac{\Dbar_{1,1,1,1}(U,V)}{UV}\,.
}
Note that the position space expression is identical to that of the $\mathcal{N}=2$ case, corresponding to the 1-loop box diagram, but the Mellin space answer looks different because the definition of Mellin space differs.
One can compute the weak coupling expansion of the planar scaling dimension of the lowest dimension single-trace scalar operator, i.e.\ the Konishi, to essentially any order using the quantum spectral curve (QSC)~\cite{Gromov:2015dfa}. This has been carried out to 11-loop order in~\cite{Marboe:2018ugv}, of which we report for simplicity only the result to 4 loops:
\begin{equation}\label{weakData4}
\Delta_{K}^{\cN = 4} = 2 + \frac{12}{(4\pi)^{2}}\lambda - \frac{48}{(4\pi)^{4}}\lambda^{2} + \frac{366}{(4\pi)^{6}} \lambda^{3} + \frac{576 \zeta_{3} - 1440 \zeta_{5} - 2496}{(4\pi)^{8}}\lambda^{4} + O(\lambda^{5}) \ed
\end{equation}
In fact, as discussed earlier, the scaling dimension can be computed at finite $\lambda$ using the QSC~\cite{Gromov:2013pga}. Nonetheless, it will be interesting to compare our bounds to a strong coupling expansion, to see when we are probing the strong coupling regime.
The leading term at strong coupling (i.e.\ $\lambda\to\infty$) for the correlator is the tree-level graviton exchange, which takes the form~\cite{Rastelli:2016nze,Arutyunov:2000py}
\es{MstrongN4}{
    \cM_{\rm strong}(s,t)=\frac{1}{(\tfrac s2-3)^2(\tfrac t2-3)^2(\tfrac{\tilde u}{2}-3)^2}\,,\qquad \cT_{\rm strong}(U,V)=-\Dbar_{2,4,2,2}(U,V)\,.
}
As in the $\mathcal{N}=2$ case, the planar CFT data of single-trace operators at strong coupling can be read off from the AdS Virasoro-Shapiro amplitude, which for the Konishi gives
\es{strongData4}{
    \Delta_K^{\mathcal{N}=4}=2\lambda^{1/4}-2+2\lambda^{-1/4}+\left(\frac12-3\zeta_3\right)\lambda^{-3/4}+O(\lambda^{-5/4})\,.
}
The finite $\lambda$ results from the QSC have been used to numerically read off further corrections to order $\lambda^{-9/4}$ in~\cite{Hegedus:2016eop}, but we will not make use of them.

\subsection{Dispersion relations}
\label{dispersion}
We now derive various dispersion relations that constrain the planar correlator. We start by deriving new results for the $\mathcal{N}=2$ case, before reviewing known results for $\mathcal{N}=4$ from~\cite{Caron-Huot:2022sdy,Caron-Huot:2024loc}.

Double-trace operators are an obstruction: they already appear in the disconnected correlator, and they enter the interacting part through $1/N$ corrections to their dimensions and OPE coefficients, which have indefinite sign.
Their contribution therefore spoils the positivity required by the numerical bootstrap.
We can project out the double-trace operators by considering the double-discontinuity $\dDisc\,T$ of the reduced correlator, following~\cite{Caron-Huot:2017vep}.
When both $z$ and $\bar z$ lie on the same cut, the double-discontinuity is defined by taking independent discontinuities in $z$ and $\bar z$~\cite{Caron-Huot:2020adz},
\es{dDiscdef}{
    \dDisc_s[T(z,\bar z)]&=\tfrac12\big(T(z_\curvearrowleft,\bar z_{\cwa})+T(z_{\cwa},\bar z_\curvearrowleft)-T(z_\curvearrowleft,\bar z_\curvearrowleft)-T(z_{\cwa},\bar z_{\cwa})\big)\quad (z,\bar z<0)\ec\\
    \dDisc_t[T(z,\bar z)]&=\tfrac12\big(T(z_\curvearrowright,\bar z_{\cwar})+T(z_{\cwar},\bar z_\curvearrowright)-T(z_\curvearrowright,\bar z_\curvearrowright)-T(z_{\cwar},\bar z_{\cwar})\big)\quad (z,\bar z>1)\ec
}
where the arrows denote the continuation of $z$ and $\bar z$ that we take around the cut starting from the region $0<z,\bar{z}<1$.
Thanks to the crossing symmetry of the correlator, the double-discontinuity in the $t$-channel contains the same information as the double-discontinuity in the $s$-channel; therefore, in the following we work only with $\dDisc\equiv\dDisc_s$.
Acting on a long block of twist $\tau=\Delta-J$, it produces the factor
\es{dDiscblock}{
    \dDisc{\!\big[(z\bar z)^{-3}g_{\tau+2,J}(z,\bar z)\big]}=2\sin^2{\!\left(\frac{\pi\tau}{2}\right)}\,(z\bar z)^{-3}g_{\tau+2,J}(z,\bar z)\ec
}
which has a double zero at the twist $\tau=4+2n$ of double-trace operators. Hence, up to order $1/N$, the double-discontinuity is saturated by single-trace operators alone. We can take the double-discontinuity of~\eqref{colorstrippedOPE} to find
\es{dDiscTcolor}{
    &\dDisc T(U,V) = \dDisc T_\text{short}(U,V)\\
    &\qquad + \frac{1}{12 k}\Big[\sum_{\tau,J^+}\lambda^2_{\mathbf{35}_v,\tau,J}\,2\sin^2(\tfrac{\pi\tau}{2})\,U^{-3}g_{\tau+2,J}+\sum_{\tau,J^-}\lambda^2_{\mathbf{28},\tau,J}\,2\sin^2(\tfrac{\pi\tau}{2})\,U^{-3}g_{\tau+2,J}\Big] \ec
}
where $J^\pm$ denote even/odd spins, and
\es{dDiscTshort}{
    \dDisc T_\text{short}(U,V)=-\frac{2}{k}\frac{1-V+V \log{V}}{V (1-V)^2}\dDisc\!\Big(\frac{1}{U}\Big)\,.
}
The sum now runs only over single-trace operators, recovering positivity.

We can then recover the full correlator from its double-discontinuity by a dispersion relation~\cite{Carmi:2019cub,Caron-Huot:2020adz}. Using $1\leftrightarrow3$ crossing to write it in terms of a single-channel double-discontinuity, we find that the $U\leftrightarrow V$, $a\leftrightarrow c$ image appears explicitly as a second term,
\es{disprel}{
    T^{abcd}(U,V)=\int\limits_{U',V'>0}\!\! dU'\,dV'\,\Big[& K(U,V;U',V')\dDisc T^{abcd}(U',V')\\& + K(V,U;U',V')\dDisc T^{cbad}(U',V')\Big]\ec
}
with the universal kernel $K$ recorded in Appendix~\ref{newDis}.

In order to apply this formula to~\eqref{dDiscTcolor} we need to discuss how the long blocks and the protected part of the correlator contribute to the dispersion relation.

Let us define the Polyakov-Regge block~\cite{Penedones:2019tng,Caron-Huot:2020adz} as the dispersive transform of a single $s$-channel block,
\es{PBpos}{
    \cP^{\mathcal{N}=2}_{\tau,J}(U,V)\equiv\int\limits_{U',V'>0}\!\! dU'\,dV'\,K(U,V;U',V')\dDisc \big[(U')^{-3}g_{\tau+2,J}(U',V')\big]\,.
}
Because of~\eqref{dDiscblock}, Polyakov-Regge blocks vanish on double-trace operators. The Mellin transform of this block can be written explicitly as a sum over descendants\footnote{We combine~\cite[Eq. 2.49]{Caron-Huot:2020adz} with the identity \( \cQ^{m,\,\cN=2}_{\Delta,J}(x-\tau-2m) = (-1)^{J} \cQ^{m,\,\cN=2}_{\Delta,J}(6-x) \).}
\es{PBmellin}{
    \cP^{\mathcal{N}=2}_{\tau,J}(s,t)=(-1)^J\sum_{m=0}^{\infty}\frac{\cQ^{m,\,\cN=2}_{\tau+J+2,J}(6-s-t)}{s-(\tau+2m)}\ec
}
where the residues are the Mack polynomials~\cite{Mack:2009mi} for the \( \cN = 2 \) correlator described in Appendix~\ref{app:mack}.

The protected part of the correlator $T_{\rm short}(U,V)$ contains only the contribution of protected multiplets and does not by itself satisfy the constraints of a conformal correlator.
We should therefore not expect the dispersive transform, which by construction returns a consistent correlator, to give $T_{\rm short}(U,V)$ back.
We notice instead that it shares its double-discontinuity with the strong coupling correlator of~\eqref{strongN2},\footnote{This is expected for every CFT dual to string theory because it is equivalent to long operators, dual to stringy states, becoming heavy and decoupling from the spectrum in the supergravity limit.}
\es{dDiscshort}{
    \dDisc T_{\rm short}(U,V)=\dDisc T_{\rm gluon}(U,V)\ec
}
and since the dispersion relation reconstructs a correlator from its double-discontinuity, and the gluon exchange diagram is a \emph{bona fide} conformal correlator, integrating~\eqref{dDiscshort} against the kernel returns the gluon exchange correlator $T_{\rm gluon}(U,V)$ of~\eqref{strongN2}.\footnote{%
    We can also check this at the level of the correlator with color indices where we find the consistent result
    \[
        \dDisc\, T_{\text{short}}^{abcd}(U,V)=\dDisc\, T_{\text{gluon}}^{abcd}(U,V)=\dDisc(\tfrac{1}{U})\Bigl[2\frac{1-V+\log(V)}{k(V-1)^2}(\mathtt{t}_4-\mathtt{t}_5)-\frac{2}{k V}\mathtt{t}_6\Bigr]\,.
    \]
}
We conclude that the color-stripped correlator for \( \mathcal{N} = 2 \) admits the following expansion in terms of single-trace data and Polyakov-Regge blocks
\begin{equation}
    \begin{aligned}
      T(U,V) &= T_{\rm gluon}(U,V) \\
      &\quad + \frac{1}{12}\Big[\sum_{\tau, J^+}\lambda^2_{\mathbf{35}_v,\tau,J}\,\cP^{\mathcal{N}=2}_{\tau,J}(U,V)+\sum_{\tau,J^-}\lambda^2_{\mathbf{28},\tau,J}\,\cP^{\mathcal{N}=2}_{\tau,J}(U,V)+(U\leftrightarrow V)\Big] \, ,
    \end{aligned}
\end{equation}
where the OPE sum runs over the single-trace operators.
Similarly, the color-stripped Mellin amplitude admits the following expansion
\begin{equation}\label{OPEMellinbox}
    \begin{aligned}
      &M(s,t) = M_{\rm gluon}(s,t)\\
      &\quad +\frac{1}{12k}\Big[\sum_{\tau, J^+}\lambda^2_{\mathbf{35}_v,\tau,J}\left(\cP^{\mathcal{N}=2}_{\tau,J}(s,t) + \cP^{\mathcal{N}=2}_{\tau,J}(t,s)\right)+\sum_{\tau,J^-}\lambda^2_{\mathbf{28},\tau,J}\left(\cP^{\mathcal{N}=2}_{\tau,J}(s,t) + \cP^{\mathcal{N}=2}_{\tau,J}(t,s)\right)\Big] \ec
    \end{aligned}
\end{equation}
in terms of the Mellin transform of Polyakov-Regge blocks~\eqref{PBmellin}. This representation is valid in the domain
\begin{equation}\label{eq:Cu-def}
    \cC_{u} = \{(s,t) : \Re(s) < 2, \ \Re(t) < 2, \ \Re(u) < 4\} \ec
\end{equation}
as explained in Appendix~\ref{app:convergence}.

The dispersive sum rules were similarly derived for $\mathcal{N}=4$ SYM in~\cite{Caron-Huot:2022sdy}, starting from the block expansion in~\eqref{block4}. The result takes a similar form, which for simplicity we report only in Mellin space
\begin{equation}\label{OPEMellinbox4}
    \mathcal{M}(s,t)=\mathcal{M}_{\rm strong}(s,t)+\sum_{\tau,J^+} \lambda^2_{\tau,J}\mathcal{P}^{\mathcal{N}=4}_{\tau,J}(s,t)\ec \qquad
    \Re(s), \Re(t), \Re(\tilde{u}) < 6 \ed
\end{equation}
Again, we find that the protected terms are captured by the strong coupling amplitude~\eqref{MstrongN4}, which is now dual to tree-level graviton exchange. The sum is over $\cN=4$ Polyakov-Regge blocks of single-trace operators:
\es{PBmellin4}{
	\cP^{\cN=4}_{\tau,J}(s,t)
    = \sum_{m=0}^\infty
    \mathcal{Q}^{m, \, \cN=4}_{\tau+J+4,J}(16 - s - t)\left[\frac{1}{s-(\tau+2m+4)}+\frac{1}{t-(\tau+2m+4)}\right] \,.
}
The residues are the Mack polynomials for the \( \cN  = 4 \) correlator and are given in Appendix~\ref{app:mack}.

\subsection{Dispersive functionals}
\label{functionals}

We now describe the set of dispersive functionals that we will use to bound planar CFT data for both the $\mathcal{N}=2$ and $\mathcal{N}=4$ theories.
All the conditions that we impose on the correlator result in a sum rule of the form
\es{generalsumrule}{
    0=\omega^{\rm prot}+\sum_{\tau,J}\lambda^2_{\tau,J}\,\omega[\tau,J]\,.
}

The first constraint we consider is crossing symmetry. The $\mathcal{N}=2$ dispersion relation~\eqref{OPEMellinbox} automatically satisfies $M(s,t)=M(t,s)$. We have, however, derived dispersion relations for both the color-stripped amplitude and the amplitude with color indices independently. We can therefore impose that the two dispersion relations are consistent with the color decomposition~\eqref{colorstripMellin}, which is an independent constraint. Imposing this consistency condition yields the crossing equation\footnote{Equivalently, one can expand the resulting correlator in flavor irreps and impose the resulting 3-vector of crossing equations, which in position space take the form
    \begin{gather}\label{crossvectorpos}
        \sum\lambda^2_{\mathbf{35}_v,\tau,J}\big(\tfrac{1}{U^3}\cP^{\mathcal{N}=2}_{\tau,J}(\tfrac1U,\tfrac VU)-\cP^{\mathcal{N}=2}_{\tau,J}(U,V)\big)-\sum\lambda^2_{\mathbf{28},\tau,J}\big(\tfrac{1}{U^3}\cP^{\mathcal{N}=2}_{\tau,J}(\tfrac1U,\tfrac VU)-\cP^{\mathcal{N}=2}_{\tau,J}(U,V)\big)=0\ec\nn\\
        \sum\lambda^2_{\mathbf{35}_v,\tau,J}\big(\tfrac{1}{U^3}\cP^{\mathcal{N}=2}_{\tau,J}(\tfrac1U,\tfrac VU)+\tfrac{1}{U^3}\cP^{\mathcal{N}=2}_{\tau,J}(\tfrac VU,\tfrac1U)-\cP^{\mathcal{N}=2}_{\tau,J}(U,V)-\cP^{\mathcal{N}=2}_{\tau,J}(V,U)\big)=0\ec\\
        \sum\lambda^2_{\mathbf{35}_v,\tau,J}\,\tfrac{1}{U^3}\cP^{\mathcal{N}=2}_{\tau,J}(\tfrac1U,\tfrac VU)+\sum\lambda^2_{\mathbf{28},\tau,J}\big(\tfrac{1}{U^3}\cP^{\mathcal{N}=2}_{\tau,J}(\tfrac VU,\tfrac1U)+\cP^{\mathcal{N}=2}_{\tau,J}(U,V)+\cP^{\mathcal{N}=2}_{\tau,J}(V,U)\big)=0\ed\nn
    \end{gather}
    It can be easily shown that the vector of crossing equations is in fact equivalent to the single equation~\eqref{crossmaster}.}
\es{crossmaster}{
    \sum_{\tau, J^+} \lambda^2_{\mathbf{35}_v,\tau,J} \Xtwo +
    \sum_{\tau, J^-} \lambda^2_{\mathbf{28},\tau,J} \Xtwo = 0 \ec \qquad
    (s,t) \in \cC_{u}
}
with \( \cC_{u} \) in~\eqref{eq:Cu-def} and
\begin{equation}\label{eq:N2-crossing-functionals}
    \Xtwo \equiv \cP^{\mathcal{N}=2}_{\tau,J}(s,t) + \cP^{\mathcal{N}=2}_{\tau,J}(t,s) - (-1)^{J}  \cP^{\mathcal{N}=2}_{\tau,J}(t,u) \, ,
\end{equation}
and we recall that \( u = 6 - s - t \).

For the \( \cN=4 \) case, \( s \leftrightarrow t \) crossing is automatically satisfied because the Polyakov-Regge blocks themselves~\eqref{PBmellin4} are invariant under this exchange. Instead, demanding \( s \leftrightarrow u \) crossing on the expansion~\eqref{OPEMellinbox4} gives the following sum rule
\es{crossmaster4}{
    0=\sum_{\tau, J^+}\lambda_{\tau,J}^2 \Xfour \,, \qquad \Re(s),\Re(t),\Re(\tilde{u})< 6 \ec
}
where
\begin{equation}
    \Xfour \equiv \mathcal{P}^{\cN=4}_{\tau,J}(s,t)-\mathcal{P}^{\cN=4}_{\tau,J}(\tilde{u},t) \ec
\end{equation}
and we recall that \( \tilde{u} = 16 - s - t \).

The crossing equations~\eqref{crossmaster} and~\eqref{crossmaster4} are of the form~\eqref{generalsumrule} with vanishing protected part $\nobreak{\omega^{\rm prot}=0}$. This is because the strong coupling correlator that appears in the dispersion relations~\eqref{OPEMellinbox} and~\eqref{OPEMellinbox4} is already crossing symmetric, unlike the unit operator that plays an analogous role in the conventional block expansion. As we will see in the next section, to numerically bound CFT data we require that functionals be normalized on a known quantity, and $\omega^{\rm prot}$ is the only option without assuming any values for OPE coefficients.

We can make progress by imposing that the correlator we are interested in is dual to a string scattering amplitude and is therefore softer in the Regge limit than the corresponding field theory amplitude~\cite{Brower:2006ea,Costa:2012cb,Penedones:2019tng}. This will result in a constraint from anti-subtracted dispersion relations~\cite{Caron-Huot:2020adz}, as first shown for the $\mathcal{N}=4$ case in~\cite{Caron-Huot:2022sdy}.

For the $\mathcal{N}=2$ case, we start with the observation that in the fixed-$u$ Regge limit
\es{}{M(s,6-u-s)=o(s^{-2})\,,}
but both the gluon exchange correlator \( M_{\rm gluon}(s,t) \) and the terms of the OPE sum in~\eqref{OPEMellinbox},
\begin{equation}
    \cP_{\tau,J}^{\cN=2}(s,t) +  \cP_{\tau,J}^{\cN=2}(t,s) = \sum_{m=0}^{\infty} \cQ^{m,\,\cN=2}_{\tau+J+2,J}(u)\left[\frac{1}{s-(\tau+2m+2)} + \frac{1}{6-s-u-(\tau+2m+2)}\right],
\end{equation}
are \( O(s^{-2}) \). This implies the sum rule
\es{Rusum}{
    0=\Ruprot +\sum_{\tau, J^+}\lambda^2_{\mathbf{35}_v,\tau,J} \Ru +\sum_{\tau, J^-}\lambda^2_{\mathbf{28},\tau,J} \Ru \ec \quad 2 < \Re(u) < 4 \ec
}
where
\es{Ru}{
    \Ru \equiv (-1)^J\sum_{m=0}^{\infty}\frac{u-6+4m+2\tau}{u-2}\,\cQ^{m, \, \cN=2}_{\tau+J+2,J}(u)\,, \qquad \Ruprot = \frac{96}{u-2} \, ,
}
and we divided everything by $u-2$ for future convenience. This functional $R^u$ oscillates with spin due to the $(-1)^J$ and is also dominant in the large-twist limit, so it is ill-suited to the numerical bootstrap, as discussed in a different context in~\cite{Lin:2015wcg,Chester:2021aun}.\footnote{In particular, if we demand that the functional be positive as needed for the numerical bootstrap, then the oscillations will set the functional to zero, making it trivial.} To avoid this problem, we derived a different dispersion relation from the fixed-$t$ Regge limit. In this limit, the amplitude behaves as $M(s,t)=o(s^{-1})$~\cite{Alday:2024yax}, while both the gluon amplitude and the individual Polyakov-Regge blocks are only $O(s^{-1})$. The leading behavior must therefore cancel in the sum. Rewriting the expansion~\eqref{OPEMellinbox} with the help of crossing~\eqref{crossmaster} as
\begin{equation}\label{eq:OPEMellinbox-cross}
    M(s,t) = M_{\rm gluon}(s,t) + \frac{1}{12k}\Big[\sum_{\tau, J^+}\lambda^2_{\mathbf{35}_v,\tau,J}\,\cP^{\mathcal{N}=2}_{\tau,J}(s,u)-\sum_{\tau,J^-}\lambda^2_{\mathbf{28},\tau,J}\,\cP^{\mathcal{N}=2}_{\tau,J}(s,u)\Big] \, ,
\end{equation}
and extracting the coefficient of $s^{-1}$, we obtain a new sum rule
\es{Rtsum}{
    0= \Rtprot +\sum_{\tau, J^+}\lambda^2_{\mathbf{35}_v,\tau,J} \Rt +\sum_{\tau, J^-}\lambda^2_{\mathbf{28},\tau,J} \Rt \ec \quad 0 < \Re(t) < 2\ec
}
where
\es{Rt}{
   \Rt =\sum_{m=0}^{\infty}\cQ^{m, \, \cN=2}_{\tau+J+2,J}(t) \,, \qquad \Rtprot = -\frac{96}{t-2} \, .
}
As advertised, there are no oscillations in this sum rule. It is the second constraint that we will impose, in addition to the crossing equation~\eqref{crossmaster}.

For $\cN=4$ SYM, the soft Regge behavior was imposed following a procedure involving the residues of the Mellin amplitude~\cite{Caron-Huot:2020adz, Caron-Huot:2022sdy}, which resulted in the sum rule
\es{ReggeN4}{
    0 = \Btprot + \sum_{\tau, J^+} \lambda_{\tau, J}^2 \Bt \ec \qquad
    4 < \Re(t) < 6 \ec
}
where
\es{Bmellin}{
   \Bt = \sum_{m=0}^\infty \frac{2\tau+4m+2-t}{t-6} \mathcal{Q}^{m,\,\cN=4}_{\tau+4,J}(10-t) \ec \qquad
   \Btprot = \frac{8}{(t-6)(t-4)} \ed
}
We find that the same sum rule can be derived following the same procedure that we used for the $\mathcal{N}=2$ case.
It suffices to consider the fixed-$u$ Regge limit of~\eqref{OPEMellinbox4} and impose $\nobreak{\cM(s,t)=o(s^{-2})}$, while both $\cM_{\rm strong}$ and the individual blocks $\cP^{\cN=4}_{\tau,J}(s,t)$ are $O(s^{-2})$. Requiring the $s^{-2}$ terms to cancel reproduces exactly~\eqref{ReggeN4}.

The functionals $\widetilde{R}$, $R$, and $B$ admit corresponding functionals in position space. The authors of~\cite{Caron-Huot:2020adz, Caron-Huot:2022sdy} introduced
\es{Bposition}{
    \Bv =
    \int_{v}^{\infty}\dif{v'}\int_{0}^{(\sqrt{v'}-\sqrt{v})^2}\dif{u'}\frac{v'-u'}{\pi^2 v \sqrt{\lambda(v,u',v')}}\dDisc[g^{\cN=4}_{\tau,J}(u',v')]\,,
}
where $\lambda(a,b,c)=  a^2 + b^2 + c^2 - 2 (a b + b c + a c)$ is the Källén lambda function, and found that it is related to~\eqref{Bmellin} via a one-dimensional Mellin transform
\es{}{
    \Bv =\frac{1}{2}\int \frac{\dif{t}}{4 \pi i} v^{\frac{t}{2}-4}\Gamma\big(4-\tfrac{t}{2}\big)^2\, \Gamma\big(\tfrac{t}{2}-1\big)^2 \Bt \ed
}
Analogously, we can define the position space version of the functionals~\eqref{Ru} and~\eqref{Rt} as an integral of the double-discontinuity of the $\cN=2$ superconformal block
\begin{align}
  \Ruv &= \int_{v}^{\infty}\!\dif{v'}\!\int_{0}^{(\sqrt{v'}-\sqrt{v})^2} \!\dif{u'}\frac{u'-v'}{\pi^2 v \sqrt{\lambda(u',v',v)}}\dDisc[g^{\mathcal{N}=2}_{\tau,J}(u',v')]\,, \\
  \Rtv &= (-1)^J\int_{v}^{\infty}\!\dif{v'}\!\int_{0}^{(\sqrt{v'}-\sqrt{v})^2} \!\dif{u'}\frac{1}{\pi^2 \sqrt{\lambda(u',v',v)}}\dDisc [g^{\mathcal{N}=2}_{\tau,J}(u',v')]\,. \label{posRt}
\end{align}
We find that they are related to the functionals introduced in Mellin space via
\begin{align}
  \Ruv &= \frac{1}{2}\int \frac{\dif{u}}{4\pi i}\, v^{-u/2} \Gamma\big(\tfrac{u}{2}\big)^2 \, \Gamma\big(2-\tfrac{u}{2}\big)^2 \, \Ru \ec \label{Ruposmellin} \\
  \Rtv &=\frac{1}{2}\int \frac{\dif{t}}{4\pi i}\, v^{-t/2} \Gamma\big(\tfrac{t}{2}\big)^2 \, \Gamma\big(2-\tfrac{t}{2}\big)^2 \, \Rt \ed \label{Rtposmellin}
\end{align}
We thus confirm that the $B$ functional of~\cite{Caron-Huot:2022sdy, Caron-Huot:2020adz} is equivalent to our $\widetilde{R}$ functional as can be seen from the equivalent definitions in position space. We prove~\eqref{Rtposmellin} explicitly in Appendix~\ref{app:eval}. The derivation of~\eqref{Ruposmellin} follows the same steps and we do not repeat it.

To improve the numerical results in~\cite{Caron-Huot:2022sdy, Caron-Huot:2024loc}, especially in the weak coupling regime, some projection functionals derived from the $B$ functional were used, called $\Phi$ and $\Psi$. We show that we cannot construct the analogue of the $\Phi$ functional in $\cN=2$, and instead construct a simpler version of the $\Psi$ functional that we can use in the numerical bootstrap.

Let us consider the $R$ functional in the $\tau \to 2$ limit,
\es{}{R_t[\tau,J] = a^{\mathcal{N}=2}_J(t) + (\tau-2) b^{\mathcal{N}=2}_J(t) + O((\tau-2)^2)\,,}
where $a^{\cN=2}_J$ is given by the $\tau \to 2$ limit of the Mack polynomials $\mathcal{Q}^{0,\,\cN=2}_{\Delta+2,J}(t)$ and reads
\es{}{a^{\cN=2}_J(t) = \frac{3 (-4)^{J+1} \left(\frac{5}{2}\right)_J \, {}_3F_2\left(-J,J+3,\frac{t}{2};2,2;1\right)}{(2)_J}\ed}
We will not need to discuss the functions $b_J$.
The functions $a^{\cN=2}_J(t)$ are a complete and orthogonal basis for the functions of $t$ on the axis $t=1+ i y$; in fact, it is easy to check that
\es{}{ c_n \int \frac{\dif{t}}{2 \pi i} \frac{(t-2)^2}{\sin(\pi t/2)^2} \, a^{\cN=2}_n(t) a^{\cN=2}_m(t) = \delta_{n,m}\ec}
with
\es{aJ-orth-normalization}{
c_n= \frac{\pi ^3 (-1)^n 4^{-2 n-5} (n+1) \Gamma (n+2) \Gamma (n+3)}{\Gamma \left(n+\frac{3}{2}\right) \Gamma
   \left(n+\frac{5}{2}\right)}\ed}
To define the analogue of the $\Phi$ functional, we would introduce the kernel
\es{Phil-def}{\Phil=\int \frac{\dif{t}}{2 \pi i} \Phi_{\ell}(t) R_t[\tau,J]\ec}
and we would require that
\es{Phil-condition}{\int \frac{\dif{t}}{2 \pi i} \Phi_{\ell}(t) a^{\cN=2}_J(t)=0 \ec \quad \forall J\geq 0\ec}
but since $\{a^{\cN=2}_J\}$ is a complete basis there is no such non-zero kernel.

In the $\mathcal{N}=4$ case~\cite{Caron-Huot:2022sdy}, the \( \Phi \) functional is defined by~\eqref{Phil-def} with \( R_{t} \to B_{t} \), with the kernel \( \Phi_{\ell}(t) \) fixed by imposing~\eqref{Phil-condition} only for even $J$. The kernel is then allowed to be a function of $t$ odd under $t\to 10-t$. Integrating~\eqref{ReggeN4} against \( \Phi_{\ell}(t) \) gives a sum rule\footnote{%
    See~\cite[App.\ B]{Caron-Huot:2022sdy} for a detailed expression of the \( \Phil \) functionals and their efficient evaluation. Note that \( {\Phil\vert_{\text{here}}} = {\Phi_{\ell,\ell+2}[\tau,J]\vert_{\text{there}}} \).
}
\begin{equation}
    0 = \sum_{\tau,J^{+}} \lambda_{\tau,J}^{2} \Phil \ec \qquad \ell = 0, 2, 4,\ldots \ec
\end{equation}
with \( \Philprot = 0 \) since \( \Btprot \) is an even function under \( t \to 10-t \).

Let us define the $\Psi$ functional for $\mathcal{N}=2$ as
\es{}{\Psil = \int \frac{\dif{t}}{2 \pi i} \Psi_{\ell}(t) R_t[\tau,J]\ec}
requiring $\Psi_{\ell}[2,J]=\delta_{\ell,J}$, or equivalently
\es{}{\int \frac{\dif{t}}{2 \pi i} \Psi_{\ell}(t) a^{\cN=2}_J(t)=\delta_{\ell,J}\ed}
Using the orthogonality of the polynomials $a^{\cN=2}_J$, it is easy to see that the only solution is
\es{}{\Psi_\ell(t)=c_\ell  \frac{(t-2)^2}{\sin(\pi t/2)^2} a^{\cN=2}_\ell(t)\ec}
with \( c_{\ell} \) given in~\eqref{aJ-orth-normalization}.
Integrating~\eqref{Rtsum} against this kernel gives the sum rule
\begin{equation}
    0 = \Psilprot +\sum_{\tau, J^+}\lambda^2_{\mathbf{35}_v,\tau,J} \Psil +\sum_{\tau, J^-}\lambda^2_{\mathbf{28},\tau,J} \Psil \ec \qquad \ell = 0,1,2,\ldots \ec
\end{equation}
where
\es{}{\Psilprot = \int \frac{\dif{t}}{2\pi i}\Psi_{\ell}(t)\Rtprot=-\frac{3 \times 4^{1-\ell}(2)_\ell}{(3/2)_\ell}\ed}
Note that this corresponds to the OPE coefficients at leading order in weak coupling.

\subsection{Localization functionals}
Finally, we consider constraints from supersymmetric localization that allow us to inject the 't Hooft coupling $\lambda$ into the bootstrap. For the $\mathcal{N}=2$ theory, we have one integrated constraint in the planar limit that acts on the Mellin amplitude as~\cite{Chester:2022sqb,Behan:2023fqq,Alday:2024yax}
\es{IntKernel2}{
    \mathcal{I}[M]=-\int\frac{\dif{s}\,\dif{t}}{(4\pi i)^{2}}M(s,t)\,\Gamma\big(\tfrac s2\big)\,\Gamma\big(2-\tfrac s2\big)\,\Gamma\big(\tfrac t2\big)\,\Gamma\big(2-\tfrac t2\big)\,\Gamma\big(\tfrac u2\big)\,\Gamma\big(2-\tfrac u2\big)\\[0.5em]
    \times\left[\frac{H_{s/2-1}+H_{1-s/2}}{(t-2)(u-2)}+\frac{H_{t/2-1}+H_{1-t/2}}{(s-2)(u-2)}+\frac{H_{u/2-1}+H_{1-u/2}}{(t-2)(s-2)}\right]\,.
}
This constraint can be evaluated using supersymmetric localization applied to mass derivatives of the sphere free energy, which in the planar limit takes the form~\cite{Chester:2022sqb,Behan:2023fqq,Alday:2024yax}
\es{locN2}{
    -3\zeta_3 + \frac{4\pi}{\sqrt\lambda}\int_0^\infty\frac{w^2\,\dif{\omega}}{(\sinh \omega)^2} \,J_1\!\big(\tfrac{\sqrt\lambda \omega}{\pi}\big) =\,k(\mathcal{I}[M(s,t)]+\mathcal{I}[M(t,u)]+\mathcal{I}[M(u,s)])\,.
}
We can act with this constraint on the dispersion relation~\eqref{OPEMellinbox} and use crossing symmetry to get
\begin{equation}\label{opeconstraint}
    0 = \Itwoprot + \sum_{\tau,J^+}\lambda^2_{\mathbf{35}_v,\tau,J} \Itwo +  \sum_{\tau,J^-}\lambda^2_{\mathbf{28},\tau,J} \Itwo
\end{equation}
where
\begin{equation}
    \Itwo \equiv \mathcal{I}[\cP^{\mathcal{N}=2}_{\tau,J}] \ec \qquad
    \Itwoprot \equiv - 2 \frac{4\pi}{\sqrt\lambda}\int_0^\infty\frac{\omega^2\,\dif{\omega}}{(\sinh \omega)^2} \, J_1\!\big(\tfrac{\sqrt\lambda \omega}{\pi}\big) \ed
\end{equation}
We note that the free-theory and strong coupling contributions exactly cancel, so one is left with a constraint on the single-trace data alone.
We discuss how to evaluate $\Itwo$ in Appendix~\ref{app:eval}. In Appendix~\ref{weak}, we verify that this integrated constraint is satisfied by the weak coupling~\eqref{weakN2} and strong coupling~\eqref{strongN2} results.

In the $\mathcal{N}=4$ case there are two independent integrated constraints, which act on the Mellin amplitude as~\cite{Binder:2019jwn,Chester:2020dja,Caron-Huot:2024loc}
\es{I2N4}{
    \mathcal{I}_2[\mathcal{M}]=-\frac12\int\frac{\dif{s}\,\dif{t}}{(4\pi i)^{2}}\,\cM(s,t) \Gamma(\tfrac s2-2)\Gamma(4-\tfrac s2)\Gamma(\tfrac t2-2)\Gamma(4-\tfrac t2)\Gamma(\tfrac u2-2)\Gamma(4-\tfrac u2)  \ec
}
\es{I4N4}{
    \mathcal{I}_4[\mathcal{M}]=&-48\int\frac{\dif{s}\,\dif{t}}{(4\pi i)^{2}}\,\cM(s,t)\Gamma(\tfrac s2-2)\Gamma(4-\tfrac s2)\Gamma(\tfrac t2-2)\Gamma(4-\tfrac t2)\Gamma(\tfrac u2-2)\Gamma(4-\tfrac u2) \\
    &\qquad\quad\times\left[\frac{2(u-5)}{(s-6)(t-6)}+\frac{t-s}{u-6}\big(H_{s/2-3}+H_{3-s/2}\big)\right]\,.
}
In the planar limit, localization fixes their values at any coupling as integrals over Bessel functions~\cite{Binder:2019jwn,Chester:2020dja},
\es{IBesselN4}{
    &I_2(\lambda)=\int_0^\infty\frac{\omega\,\dif{\omega}}{(\sinh\omega)^2}\Big( J_1(\tfrac{\omega\sqrt{\lambda}}{\pi})^2-J_2(\tfrac{\omega\sqrt{\lambda}}{\pi})^2\Big)\ec\\
    &I_4(\lambda)=\,48\zeta_3-\frac{128\pi^2}{\lambda}\int_0^\infty\frac{\omega\,\dif{\omega}}{(\sinh\omega)^2}J_1(\tfrac{\omega\sqrt{\lambda}}{\pi})^2\\
    &-\frac{384\pi}{\sqrt{\lambda}} \int_0^\infty\frac{\omega\,\dif{\omega}\,J_1(\tfrac{\omega\sqrt{\lambda}}{\pi})}{(\sinh\omega)^2} \int_0^\infty\frac{\omega'\,\dif{\omega'}\,J_1(\tfrac{\omega'\sqrt{\lambda}}{\pi})}{(\sinh\omega')^2}
    \left(\frac{\omega\,J_0(\tfrac{\omega\sqrt{\lambda}}{\pi})J_1(\tfrac{\omega'\sqrt{\lambda}}{\pi})-(\omega\leftrightarrow \omega')}{\omega'^2-\omega^2}\right)\ec
}
so that inserting the Polyakov-Regge expansion yields two sum rules
\es{}{0=\Ifourpprot +\sum_{(\tau,J)}\lambda^2_{\tau,J}\, \Ifourp \, \quad (p=2,4)\,,}
with
\es{IprotN4}{
    \Ifourp \equiv \mathcal{I}_{p}\left[\cP^{\cN=4}_{\tau, J}\right] \ec \qquad
    \Ifourtwoprot \equiv\frac14-I_2(\lambda)\ec\qquad
    \Ifourfourprot\equiv 24(2\zeta_3-1)-I_4(\lambda)\ec
}
where the two constants are the values of~\eqref{I2N4} and~\eqref{I4N4} evaluated on the supergravity amplitude $\cM_{\rm strong}$. In the strong coupling limit $\lambda\to\infty$ the normalization of the sum rule vanishes, $\Ifourpprot\to0$, as it must, since all long single-trace operators become heavy and decouple. The weak coupling limit was also verified in~\cite{Wen:2022oky}.
The efficient evaluation of $\Ifourp$ was discussed in~\cite{Caron-Huot:2024loc}.

\section{Numerical Bootstrap}
\label{numBoot}

\subsection{Setup}\label{numsetup}

In this section we combine the various sum rules described in Section~\ref{setup} to bound the scaling dimension of the lightest single-trace scalar operator in the \( \cN = 4 \) and \( \cN = 2 \) theories.

The sum rules take the form~\eqref{generalsumrule} and use a combination of different dispersive functionals \( \omega_{k} \), where \( k \) labels the different sum rules. We summarize in Table~\ref{tab:functionals-menu} the menu of functionals that we will use and the meaning of \( k \) in each case.

For the case where \( k = (s,t) \) or \( k = t \), we sample the domain where the corresponding sum rule holds, see Appendix~\ref{app:convergence} for more details. The crossing sum rule is labeled by pairs \( (s,t) \) that we sample as follows
\begin{align}
  \cN=4 \quad \colon \quad (s, t) &\in \left\{\big(6-\tfrac{i}{n_{X}+1}, 6-\tfrac{2 j}{n_{X}+1}\big) \colon i = 1,\ldots, n_{X} \ec \ j = 1,\ldots,n_{X}-i\right\} \ec   \label{stListN4} \\
  \cN=2 \quad \colon \quad (s, t) &\in \left\{\big(2-\tfrac{i}{n_{X}},2-\tfrac{j}{n_{X}}\big) \colon i = 1,\ldots,2n_{X} \ec \ j = 1,\ldots,2n_{X}-i-1\right\} \ec \label{stListN2}
\end{align}
where \( n_{X} \) is an integer parameter.
For the anti-subtracted sum rule labeled by \( t \), we sample the domain as
\begin{align}
    \cN=4 \quad \colon \quad t &\in \left\{4 + \tfrac{2 i}{n_{A} + 1} \colon i = 1,\ldots,n_{A} \right\} \label{tListN4} \\
    \cN=2 \quad \colon \quad t &\in \left\{\tfrac{1}{2}+\tfrac{i-1}{n_{A}-1}: i=1,\dots,n_A\right\} \label{tListN2}
\end{align}
where \( n_{A} \) is an integer parameter, while the weak coupling functionals are imposed for \( \ell = 0,\ldots, \ell_{\rm max} \) (restricted to even \( \ell \) in the \( \cN = 4 \) case). The total number of functionals thus scales as
\begin{equation}\label{eq:total-number-functionals}
    \begin{aligned}
      \cN=4 \quad \colon \quad N_{\rm tot} &= \tfrac{1}{2}n_{X}(n_{X}-1) + n_{A} + \big(\tfrac{1}{2}\ell_{\rm max} + 1\big) + 2 \\
      \cN=2 \quad \colon \quad N_{\rm tot} &= (2n_{X}-1)(n_{X}-1) + n_{A} + (\ell_{\rm max} + 1) + 1
    \end{aligned}
\end{equation}

The numerical optimization used to bound scaling dimensions of operators is analogous to the standard numerical bootstrap algorithm, see~\cite{Poland:2018epd,Chester:2019wfx} for a review. For a fixed set of functionals \( (n_{X}, n_{A}, \ell_{\rm max}) \) and for a fixed value of the coupling \( \lambda \) we look for a dual functional \( \alpha_{k} = (\alpha_{1},\ldots,\alpha_{N_{\rm tot}}) \) satisfying the following constraints
\begin{equation}\label{eq:gap-maximization-sdp}
    \begin{aligned}
      \sum_{k} \alpha_{k} \omega_{k}^{\rm prot} &= 1 \ec \\
      \sum_{k} \alpha_{k} \omega_{k}[\tau, 0] &\geq 0 \ec \qquad && \tau \geq \tau_{\rm gap} \ec\\
      \sum_{k} \alpha_{k} \omega_{k}[\tau, J] &\geq 0  \ec \qquad && \tau \geq 2 \ec \ J > 0 \ed
    \end{aligned}
\end{equation}
If the functional exists, the positivity and normalization constraints are incompatible with the sum rules~\eqref{generalsumrule} and thus exclude the existence of a theory with a gap of \( \tau_{\rm gap} \) in the spin \( J = 0 \) single-trace sector. If we cannot find the functional, the assumed spectrum is compatible with the sum rules we considered. By bisecting in \( \tau_{\rm gap} \) we find the maximal allowed gap for the scalar single-trace sector, i.e.\ we solve the \emph{gap maximization problem}. Recall that \( \omega_{k}^{\rm prot} \) depends on the coupling \( \lambda \) through \( \Ifourtwoprot \) and \( \Ifourfourprot \) in the \( \cN=4 \) case, and through \( \Itwoprot \) for the \( \cN=2 \) case. By solving the gap maximization problem for many values of \( \lambda \) we obtain a bound on the scaling dimension of the lightest single-trace scalar operator as a function of the coupling.
Note that \( J \) in~\eqref{eq:gap-maximization-sdp} is restricted to be an even integer in the \( \cN = 4 \) case.

\begin{table}[t]
    \centering
    \renewcommand{\arraystretch}{1.5}
    \sbox0{
      \begin{tabular}[t]{ccc}\toprule
        functional & protected & \( k \) \\ \midrule
        \( \Xfour \) & \( 0 \) & \( (s,t) \) \\
        \( \Bt \) & \( \Btprot \) & \( t \) \\
        \( \Phi_{\ell}[\tau, J] \) & \( 0 \) & \( \ell \) \\
        \( \Ifourtwo \) & \( \Ifourtwoprot \) & -- \\
        \( \Ifourfour \) & \( \Ifourfourprot \) & -- \\ \bottomrule
      \end{tabular}}%
    \begin{subfigure}[t]{0.45\linewidth}
        \centering
        \begin{minipage}[t][\dimexpr\ht0+\dp0\relax][t]{\linewidth}
            \centering\usebox0
        \end{minipage}
        \caption{\( \cN = 4 \)}
    \end{subfigure}
    \begin{subfigure}[t]{0.45\linewidth}
        \centering
        \begin{minipage}[t][\dimexpr\ht0+\dp0\relax][t]{\linewidth}
            \centering
            \begin{tabular}[t]{ccc}\toprule
              functional & protected & \( k \) \\ \midrule
              \( \Xtwo \) & \( 0 \) & \( (s,t) \) \\
              \( \Rt \) & \( \Rtprot \) & \( t \) \\
              \( \Psi_{\ell}[\tau, J] \) & \( \Psi_{\ell}^{\rm prot} \) & \( \ell \) \\
              \( \Itwo \) & \( \Itwoprot \) & -- \\ \bottomrule
            \end{tabular}
        \end{minipage}
        \caption{\( \cN = 2 \)}
    \end{subfigure}
    \caption{Menu of the dispersive functionals used in this work, obtained from crossing, anti-subtracted dispersion relations and localization constraints.\label{tab:functionals-menu}}
\end{table}

To solve the constraints~\eqref{eq:gap-maximization-sdp} numerically we need to perform two further truncations since we cannot deal with the continuous range of \( \tau \)\footnote{%
    In the usual conformal bootstrap, the functionals \( \omega[\tau,J] \) are proportional to derivatives of the conformal blocks. The latter can be approximated by rational functions in \( \Delta = \tau + J \) with a positive denominator, so that the analogue of~\eqref{eq:gap-maximization-sdp} can be solved for continuous \( \Delta \) using a polynomial matrix program. It would be interesting to explore whether the functionals used in this work admit accurate polynomial interpolations in \( \tau \) in the spirit of~\cite{Chang:2025mwt}.
} and with an infinite number of spins.

For the twist, we sample the range \( \tau \in [2, 30] \) over \( \approx 620 \) values and check dependence on this grid. As described in~\cite{Caron-Huot:2024loc}, the sensitivity to the omitted large-twist values can be greatly reduced by supplementing~\eqref{eq:gap-maximization-sdp} with an additional positivity constraint that corresponds to the Regge limit of the dispersive sum rules. We postpone the discussion of this technical detail to Appendix~\ref{app:regge-limit}. With this extra constraint, our bounds are stable with respect to the sampling in twist.

For the spin, we impose the positivity constraints for \( J = 0, \ldots, J_{\rm max} \) and check the dependence of our bounds on \( J_{\rm max} \).
The bounds that we obtain can be quite sensitive to the spin truncation.
The general pattern that we observe is that the small \( \lambda \) region (weak coupling) requires a high \( J_{\rm max} \).
Moreover, increasing \( n_{X} \) and \( n_{A} \) requires a higher \( J_{\rm max} \).
If the spin truncation is not large enough, we find everything disallowed below some value of the coupling, and a normal bound above that.
The intuition behind this behavior is that the leading twist-2 trajectory has relatively small twist \( \tau \sim 2 + \# \lambda \log J + \cdots \) at weak coupling and large spin~\cite{Alday:2015eya}, and can thus give important contributions to the sum rule.\footnote{%
    Note that the dispersive sum rules have only power-law convergence~\cite{Caron-Huot:2022sdy}, contrary to the regular conformal block decomposition, which converges exponentially~\cite{Pappadopulo:2012jk}.
}
A functional that is positive only up to \( J_{\rm max} \) can violate positivity near these physical operators and invalidate the bound at weak coupling.
We postpone a more detailed discussion of the spin-truncation effects to Sections~\ref{resultsN4} and~\ref{resultsN2} for the \( \cN = 4 \) and \( \cN=2 \) cases, respectively.

After having performed these truncations, the positivity constraints~\eqref{eq:gap-maximization-sdp} become a linear program which we solve using \texttt{SDPB}~\cite{Simmons-Duffin:2015qma,Landry:2019qug}.

\subsection{Results for \( \cN = 4 \)}
\label{resultsN4}

\begin{figure}[h]
    \centering
    \includegraphics[width=\linewidth]{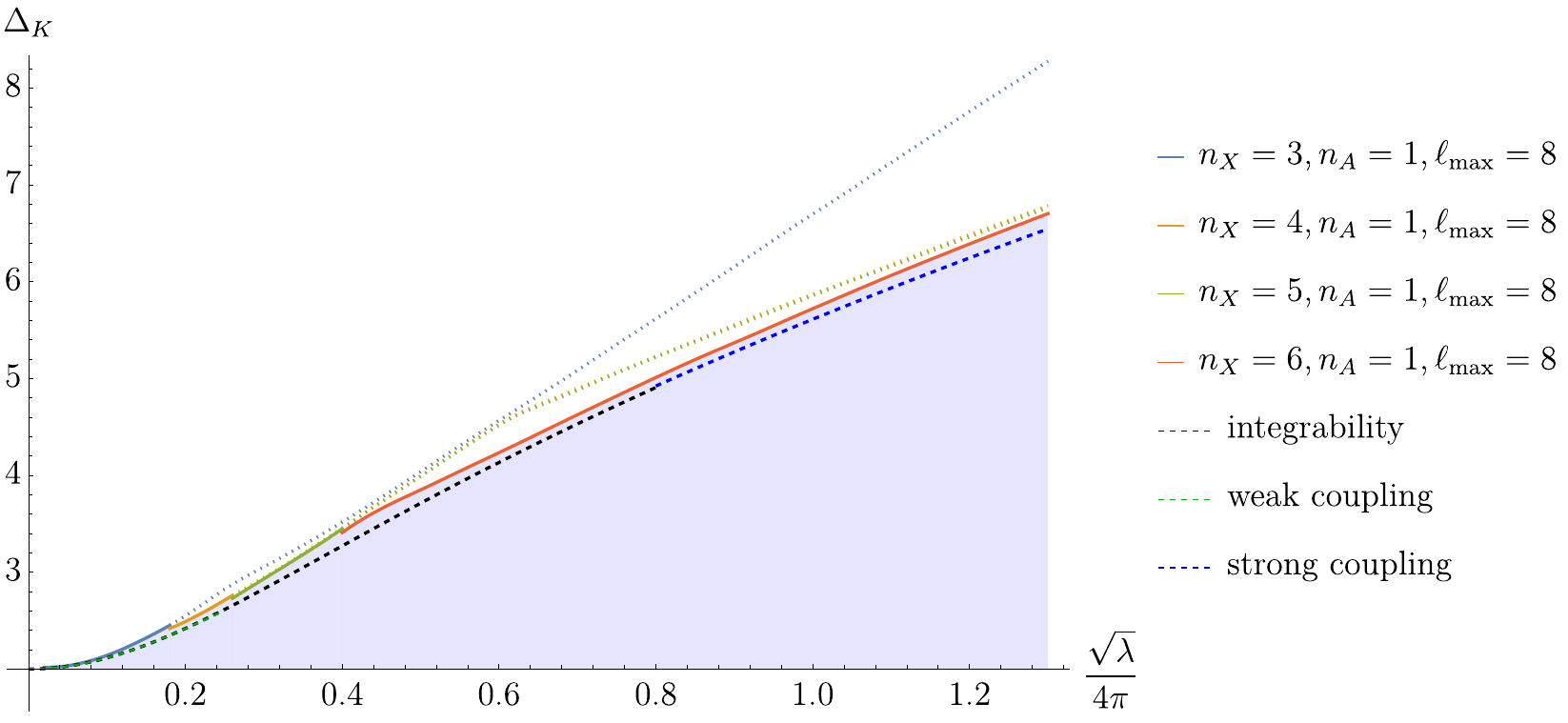}
    \caption{Bound on the scaling dimension \( \Delta_{K} \) of the lightest single-trace scalar primary in the planar \( \cN = 4 \) theory as a function of the 't Hooft coupling \( \lambda \). The shaded blue area is allowed. The thick dashed lines denote the weak coupling~\eqref{weakData4}, strong coupling~\eqref{strongData4} (including the \( \lambda^{-5/4} \) correction from~\cite{Hegedus:2016eop}) and the integrability~\cite{Gromov:2023hzc} predictions in planar \( \cN = 4 \) SYM. The bounds are obtained with combinations of the functionals in Table~\ref{tab:functionals-menu} and become stronger as we increase the number of crossing functionals \( X_{s,t} \). See the main text for a discussion on the logic for joining the various bounds as we vary \( \lambda \).\label{fig:boundN4}}
\end{figure}

In Figure~\ref{fig:boundN4} we show the best bound we have obtained on the scaling dimension \( \Delta_{K} \) of the lightest single-trace scalar, the Konishi operator, in the planar \( \cN = 4 \) theory as a function of the coupling.
The shaded light-blue area is allowed by the planar bootstrap. The dashed thick lines denote the integrability (black), weak coupling (green) and strong coupling (blue) values for \( \Delta_{K}(\lambda) \).

The bound is obtained using a combination of the functionals in Table~\ref{tab:functionals-menu} as described in the legend. For this bound we keep the anti-subtracted functionals fixed to one \( B \) functional (\( n_{A} = 1 \)) and five weak coupling \( \Phi_{\ell} \) functionals (\( \ell_{\rm max} = 8 \)). We increase the number of crossing \( X \) functionals from \( n_{X} = 3, 4, 5, 6 \) to inspect convergence.
The bound is obtained at \( J_{\rm max} = 800 \), which is the reason for the slightly unpleasant discontinuities in this bound. As anticipated in Section~\ref{numsetup}, for a fixed number of functionals, in this case for fixed \( n_{X} \), we obtain a bound only for \( \lambda \geq \lambda_{*}(J_{\rm max}, n_{X}) \). For \( \lambda < \lambda_{*}(J_{\rm max}, n_{X}) \) we find everything disallowed but we believe this is just an artifact of spin truncation. The value \( \lambda_{*}(J_{\rm max}, n_{X}) \) increases with \( n_{X} \) at fixed \( J_{\rm max} \)\footnote{We find empirically that it is almost linear in the total number of \( X \) functionals \( \tfrac{1}{2}n_{X}(n_{X}-1) \).} and decreases with \( J_{\rm max} \) at fixed \( n_{X} \). In theory, one should increase \( J_{\rm max} \) as one increases \( n_{X} \), trying to keep \( \lambda_{*} \) fixed and as small as possible. However, it becomes very expensive to numerically compute the dispersive functionals for high \( J_{\rm max} \) and we thus settled on the more pragmatic choice of ``gluing'' our bounds for different ranges of \( \lambda \). In Figure~\ref{fig:boundN4} we draw as a solid line the strongest bound we can get for that value of \( \sqrt{\lambda}/4\pi \), and as a dashed line the bound with fewer functionals, when it is non-trivial.

For \(\sqrt{\lambda}/4\pi \geq 0.5 \) we can compare bounds with \( n_{X} = 3,4,5,6 \) and observe nice convergence to both the integrability and the strong coupling predictions. Some of these bounds present some kinks which disappear as we increase the number of functionals. For \( 0.26 \leq \sqrt{\lambda}/4\pi \leq 0.4 \) only \( n_{X} = 3,4,5 \) are non-trivial and almost overlap, and it is the region where the bound is furthest from the integrability prediction. For \( 0.18 \leq \sqrt{\lambda}/4\pi \leq 0.26 \) we can compare the \( n_{X} = 3,4 \) bounds, which already show some improvement towards the integrability and weak coupling predictions. For \( 0.01 \leq \sqrt{\lambda}/4\pi \leq 0.17 \) only the \( n_{X} = 3 \) bound is available and gets extremely close to the weak coupling and integrability predictions.\footnote{Such nice agreement in this region is due mostly to the anti-subtracted and weak coupling functionals. Without those the bound would be much further at weak coupling.}
For \( \sqrt{\lambda}/4\pi \geq 2 \) we have also obtained bounds with \( n_{X} = 7,8,9 \), which show a mild improvement with respect to the \( n_{X} = 6 \) bound. For large values of \( \lambda \), however, the bound starts to deviate from the strong coupling prediction, probably because \( \Ifourpprot \to 0 \) at strong coupling and thus the bootstrap becomes less and less sensitive to the localization constraint.\footnote{%
    Note that in the absence of localization constraints there is no bound on the lightest single-trace scalar since in planar \( \cN = 4 \) SYM \( \Delta_{K} \sim 2 \lambda^{1/4} \to +\infty \) as \( \lambda \to +\infty \).
}
In any case, comparing strong coupling and integrability, we see that the strong coupling regime starts at \( \sqrt{\lambda}/4\pi \sim 0.5 \) where our bound is already quite close to the prediction.
\vspace*{-0.4cm}
\subsection{Results for \( \cN = 2 \)}
\label{resultsN2}
\vspace*{-0.3cm}
\begin{figure}[h]
    \centering
    \includegraphics[width=\linewidth]{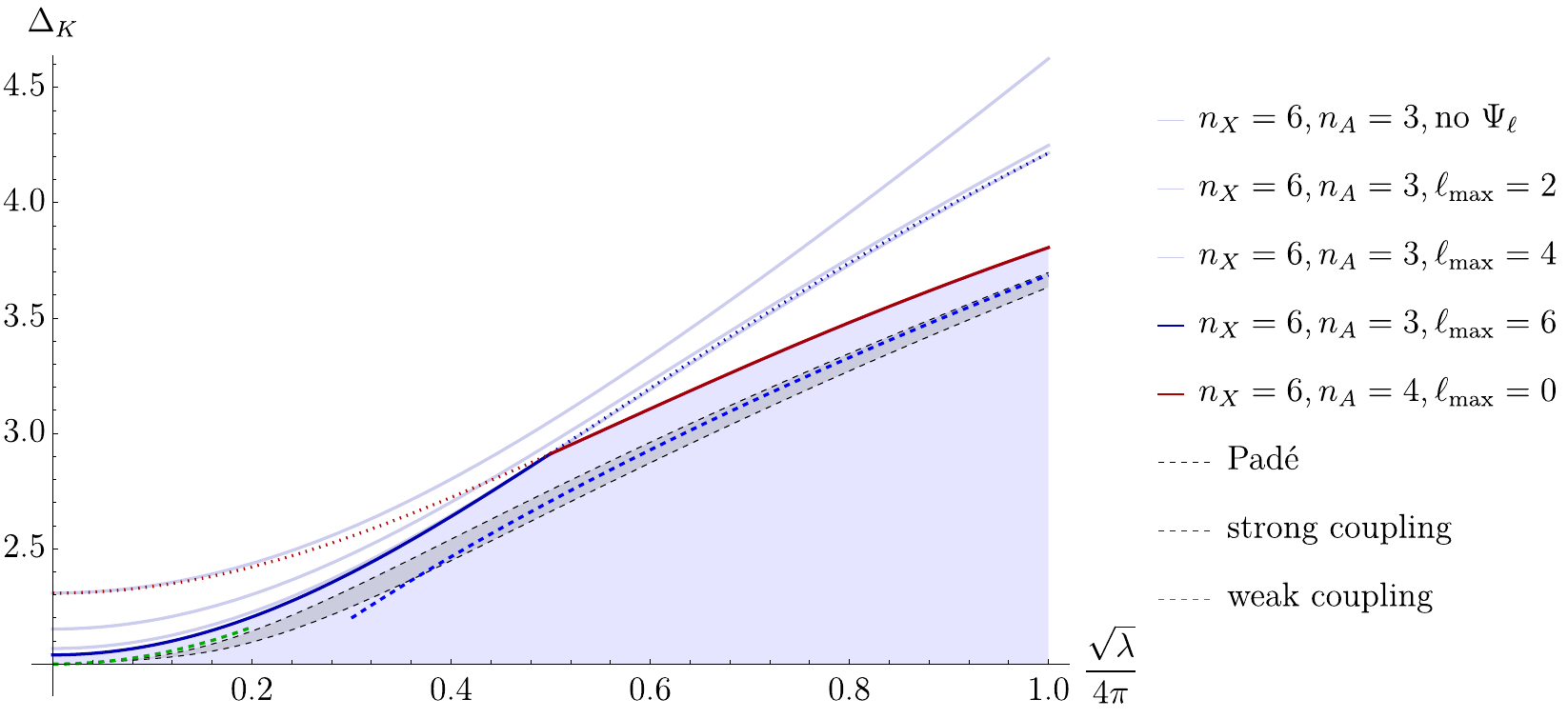}
    \vspace*{-0.8cm}
    \caption{Bound on the scaling dimension \( \Delta_{K} \) of the lightest single-trace scalar primary in the planar \( \cN = 2 \) theory as a function of the 't Hooft coupling \( \lambda \). The shaded blue area is allowed. The thick dashed lines denote the weak coupling~\eqref{weakData} and strong coupling~\eqref{strongData} prediction in the planar \( \cN = 2 \) \( U\!Sp(2N) \) theory. The gray band denotes a two-sided Padé approximation of the weak and strong coupling expansion, whose width represents a measure of the uncertainty. We went up to $\frac{\sqrt{\lambda}}{4 \pi}\sim 1.0$ because we found it numerically challenging to bound \( \Delta_{K} \) for $\frac{\sqrt{\lambda}}{4 \pi}\gtrsim 1.1$ as explained in the main text. The bounds are obtained with combinations of the functionals in Table~\ref{tab:functionals-menu}. See the main text for a discussion on the logic for joining the various bounds as we vary \( \lambda \). Notice how in this plot both the abscissa and the ordinate are more zoomed-in than in Figure~\ref{fig:boundN4} giving the illusion that the bound for $\cN=2$ is significantly worse than the bound for $\cN=4$, whereas we actually used comparable numerical effort and we got similar numerical errors in the two cases.\label{fig:boundN2}}
\end{figure}
In Figure~\ref{fig:boundN2} we show the best bound we have obtained on the scaling dimension \( \Delta_{K} \) of the lightest single-trace scalar in the planar \( \cN = 2 \) theory as a function of the coupling \(\sqrt{\lambda}/4\pi \).
The shaded light-blue area is allowed by the planar bootstrap. The dashed lines are the weak coupling~\eqref{weakData} and strong coupling~\eqref{strongData} predictions. A gray band denotes a two-sided Padé that interpolates between weak and strong coupling expansions to give us an idea of the values of \( \lambda \) for which we are probing the weak and strong coupling regimes. The thickness of the band is a measure of the systematic error coming from the choice of perturbative data included in the Padé approximation.

The bound is obtained using a combination of the functionals in Table~\ref{tab:functionals-menu} as described in the legend. For technical reasons that we discuss below, the bound is obtained with a fixed number of crossing \( X \) functionals (\( n_{X} = 6\)).

With three anti-subtracted \( R \) functionals (\( n_{A} = 3 \)), we vary the number of weak coupling \( \Psi \) functionals, taking \( \ell_{\rm max} = 0,2,4,6 \), and observe monotonic convergence of the bound, which gets quite close at weak coupling to the weak coupling prediction and to the upper edge of the Padé band for intermediate couplings. The best bound, corresponding to \( \ell_{\rm max} = 6 \), is given by the thick blue line. The bound, however, deviates significantly from the strong coupling prediction at strong coupling.

We then also report, in red on the same plot, a bound obtained with four anti-subtracted \( R \) functionals (\( n_{A} = 4 \)) and one weak coupling \( \Psi \) functional (\( \ell_{\rm max} = 0 \)). This bound is worse than the \( n_{A} = 3 \) bound at weak coupling but much more constraining and closer to the strong coupling prediction at strong coupling.

As for the \( \cN = 4 \) case, gluing bounds with different combinations of functionals is a pragmatic choice forced by problems with convergence in \( J_{\rm max} \). For the \( \cN = 2 \) setup, there is an unfortunate and more important issue that we have to face. The \( R \) functionals, defined in Section~\ref{functionals}, were introduced as an alternative sum rule to the \( \widetilde{R} \) sum rule, which was dominant and oscillatory in \( J \) for large twist \( \tau \). However, these functionals turn out to still be oscillating in \( J \) for small twist. For a fixed combination of functionals and a fixed value of \( \lambda \), we observe that there is a window in \( J_{\rm max} \) where the bound stabilizes, but after some threshold the upper bound gets worse. Investigating the associated dual functional we find that some components of the functional are, numerically, set to 0 by \texttt{SDPB}.
\begin{figure}[h]
    \centering
    \includegraphics[width=0.45 \linewidth]{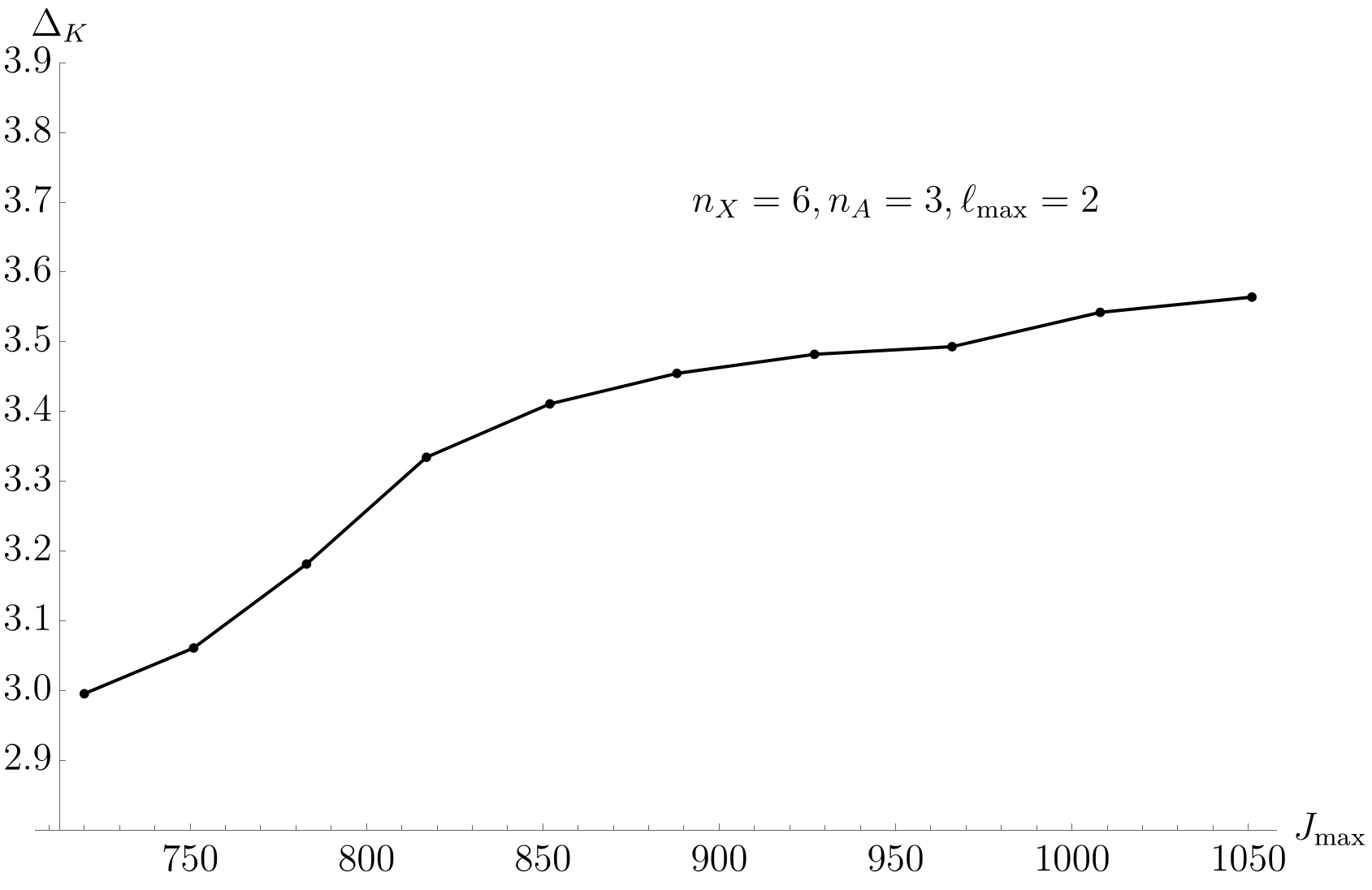}\quad \includegraphics[width=0.45 \linewidth]{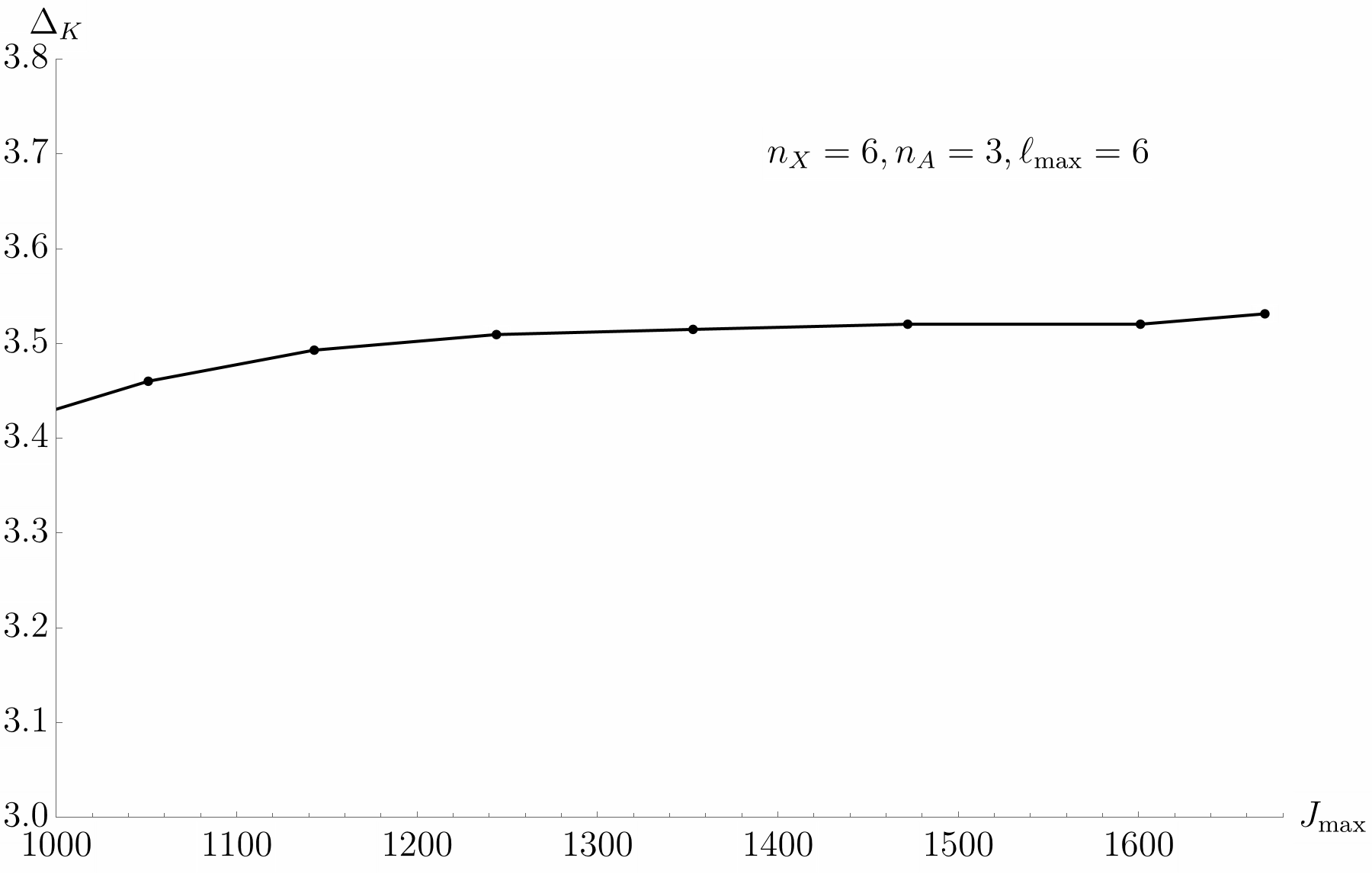}
    \caption{Bound on the scaling dimension of the Konishi operator in the \( \cN = 2 \) case as a function of the maximum spin $J_\text{max}$ for two configurations of functionals at $\sqrt{\lambda}/4 \pi=0.7$. The window of stability of the bound for $\ell_\text{max}=2$ is among the worst that we accepted, and the situation does not necessarily worsen by increasing the number of functionals, see for example $\ell_\text{max}=6$.}
\end{figure}
The width of this window depends on the coupling and the combination of functionals, and Figure~\ref{fig:boundN2} reports the configurations of functionals and values of the coupling for which we were able to identify a stable window, which we take as the bound. For weak coupling we find that we need to take larger values of $J_\text{max}$ for the bound to converge, but increasing $J_\text{max}$ further does not change the bound, hence the window where the bound is stable is very large. As we increase the coupling the window shrinks. For $n_A=3$ we find the worst behavior around $\frac{\sqrt{\lambda}}{4 \pi}=0.7$, for $n_A=4$ we find instead that the window continues shrinking increasing the coupling and disappears for  $\frac{\sqrt{\lambda}}{4 \pi}\gtrsim 1.1$.
For some configurations of functionals or for large values of \( \lambda \) the window does not even exist.

\begin{figure}[h]
    \centering
    \includegraphics[width=0.8 \linewidth]{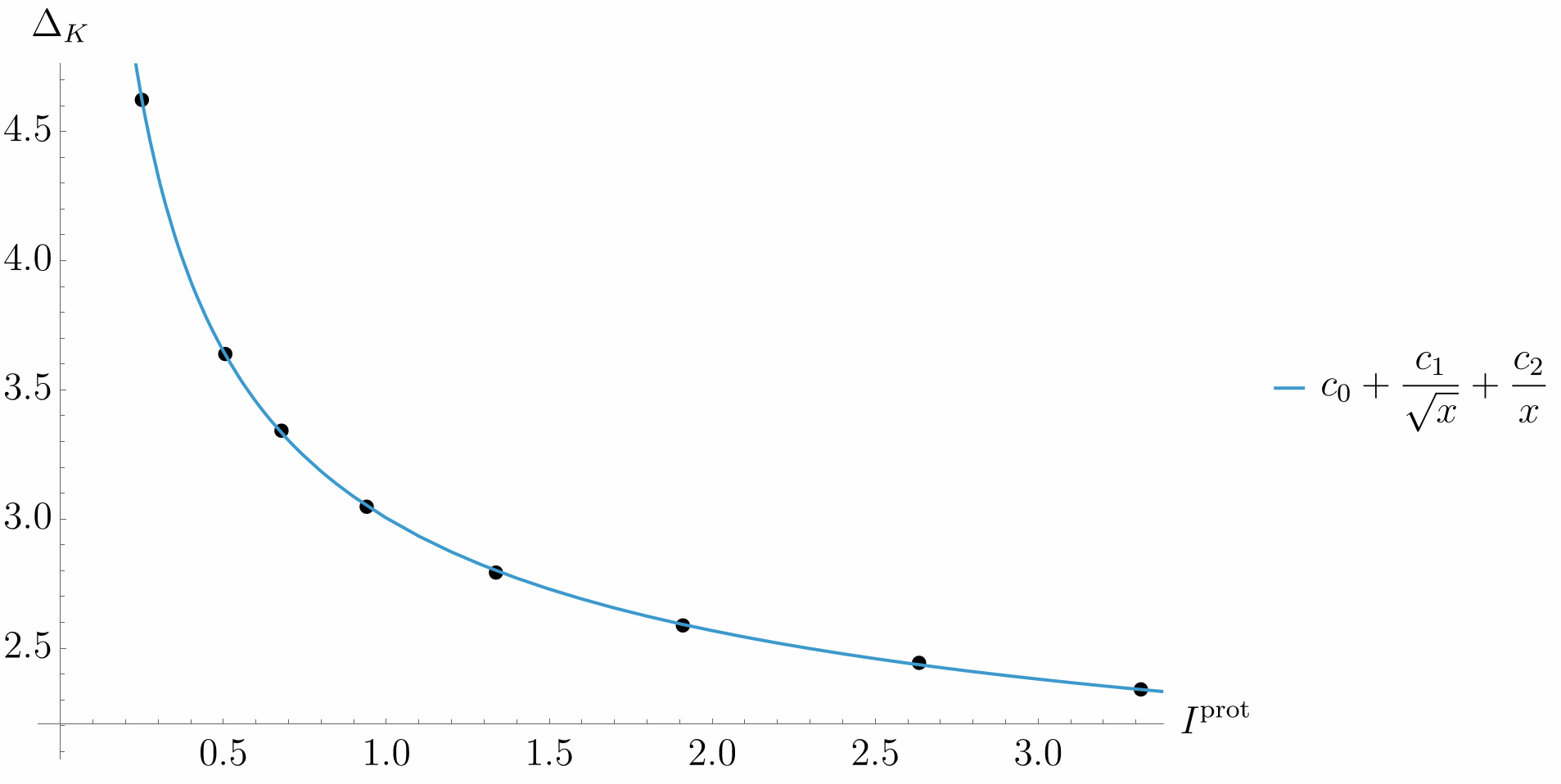}
    \caption{Upper bound on \( \Delta_{K} \) as a function of \( \Itwoprot \), the only coupling-dependent input of the \( \cN=2 \) setup, shown here for \( n_{X}=6 \), \( n_{A}=3 \). The black dots are obtained by \texttt{SDPB} and the blue curve is a simple three-parameter fit. We use this fit to plot the bound continuously in \( \lambda \) in Figure~\ref{fig:boundN2}, and as a check on the runs with a narrow stability window in \( J_{\rm max} \).
}    \label{fig:fit}
\end{figure}

In the numerical bootstrap setup for \( \cN=2 \), the only coupling dependent input is \( \Itwoprot \). Investigating the bound as a function of \( \Itwoprot \), we find that it follows a simple curve\footnote{%
    This does not work well for the \( \cN=4 \) case because we have two coupling-dependent terms in the normalization of~\eqref{eq:gap-maximization-sdp}, namely \( \Ifourtwoprot \) and \( \Ifourfourprot \).
}, as shown in Figure~\ref{fig:fit}.
We therefore plot the bound as a continuous function of the coupling by means of a simple fit in \( \Itwoprot \). This also makes us more confident about the bounds obtained with a small window of convergence as they lie on the same curve as more robust bounds obtained for $\frac{\sqrt{\lambda}}{4 \pi}\lesssim 0.5$ and $\frac{\sqrt{\lambda}}{4 \pi}\gtrsim 0.8$.

\section{Conclusion}
\label{conc}

In this paper, we showed how the numerical bootstrap combined with supersymmetric localization can be applied to holographic theories in the planar limit to compute accurate upper bounds on scaling dimensions of single-trace operators for all values of the 't Hooft coupling $\lambda$. We applied this method to two examples: the stress tensor correlator in $\mathcal{N}=4$ SYM, which is dual to closed string scattering, and the $SO(8)$ flavor multiplet correlator in a certain $\mathcal{N}=2$ CFT with $SO(8)\times SU(2)$ flavor symmetry, which is dual to open string scattering. For $\mathcal{N}=4$ SYM, we found that our bounds were close to the known result from integrability for all $\lambda$. For the $\mathcal{N}=2$ theory, for which integrability is not available, we found that our bounds were close to weak coupling results for small $\lambda$, as well as strong coupling results from the AdS Veneziano amplitude at large $\lambda$.

While our bounds in both cases were reasonably close to the predictions from other methods, we did not find the precise saturation that was observed for scaling dimensions in the finite $N$ numerical bootstrap for $\mathcal{N}=4$ SYM~\cite{Chester:2023ehi,Chester:2021aun}, or the very precise upper and lower bounds that were found for OPE coefficients in the planar bootstrap for $\mathcal{N}=4$ after inserting scaling dimensions from integrability~\cite{Caron-Huot:2024loc}. For the $\mathcal{N}=2$ theory, we might obtain better bounds by considering mixed correlators with other $\Delta=2$ external half-BPS operators, such as the chiral multiplet or the $SU(2)$ flavor multiplet. For the correlator of $SU(2)$ flavor multiplets, the localization constraint in the planar limit is expected to be the same as in $\mathcal{N}=4$ SYM,\footnote{This is because the $SU(2)$ flavor multiplet is dual to a closed string state, whose correlator is universal in the planar limit.} while the localization constraint for the mixed $SU(2)$ and $SO(8)$ correlator was derived in~\cite{Chester:2025ssu}. For the correlator of chiral multiplets, localization fixes a certain protected double-trace OPE coefficient~\cite{Baggio:2014sna,Baggio:2014ioa}, but this has no effect on the planar bootstrap. On the other hand, the mixed chiral and flavor multiplet correlators should have a non-trivial localization constraint.

The dispersion relations that we derived for the $\mathcal{N}=2$ theory should also be useful in other contexts. For instance,~\cite{Dempsey:2025yiv} combined dispersion relations with integrability, localization, and bootstrap to derive tight bounds on energy correlators for $\mathcal{N}=4$ SYM, which are related to a certain integral of the stress tensor multiplet correlator. Our planar $\mathcal{N}=2$ dispersion relations for the flavor multiplet correlator could be used for a similar study of charge correlators in this theory, though integrability results are not yet available. One could also derive similar dispersion relations for the $\mathcal{N}=2$ theory at finite $N$, which might improve the finite $N$ numerical bootstrap~\cite{Chester:2022sqb}. Indeed, we observed that the anti-subtracted $\mathcal{N}=2$ dispersion relations drastically improved the convergence of our bounds for the planar bootstrap, and so they might have a similar effect on the finite $N$ bounds, which are still far from convergence for the $N=2$ case considered in~\cite{Chester:2022sqb}.

It would be interesting to apply the planar bootstrap to other holographic theories. A prime target is 3d $\mathcal{N}=6$ $U(N)_{k}\times U(M)_{-k}$ ABJ(M)~\cite{Aharony:2008ug,Aharony:2008gk}, in the large $N,M,k$ limit for fixed $\lambda_1\equiv N/k$ and $\lambda_2\equiv M/k$. Three localization constraints have been derived for this theory in~\cite{Binder:2019mpb,Binder:2021cif}, and a fourth is expected to exist~\cite{Chester:2021gdw}. The AdS Virasoro-Shapiro amplitude has also been derived in~\cite{Chester:2024esn}, while quantum spectral curve results for some (but not all) low-lying scaling dimensions have been computed at finite $\lambda_1$ and $\lambda_2$ in~\cite{Bombardelli:2018bqz,Cavaglia:2014exa}. For the scaling dimensions that are known from integrability, one could follow the strategy of~\cite{Caron-Huot:2024loc} of inserting them into the planar bootstrap along with localization to compute upper and lower bounds on the corresponding OPE coefficients. For the unknown scaling dimensions, the strategy in this paper can give an upper bound, which could then be compared to predictions at strong coupling from the AdS Virasoro-Shapiro amplitude.

Finally, it would be fascinating if a quantum spectral curve could be derived for the open string sector of the \( 4d \) $\mathcal{N}=2$ theory we consider. Some early work on the pp-wave limit~\cite{Berenstein:2002zw}, integrability at weak coupling~\cite{Chen:2004mu,Chen:2004yf}, and classical integrability of $D7$ brane boundary conditions in $AdS_5\times S^5$~\cite{Dekel:2011ja}, all suggest this may be possible. If scaling dimensions for this theory could be derived for all $\lambda$, then they could be inserted into the planar bootstrap introduced in this work to obtain accurate upper and lower bounds for the corresponding OPE coefficients, as was previously done for $\mathcal{N}=4$ SYM in~\cite{Caron-Huot:2024loc}.

\section*{Acknowledgments}

We thank Simon Caron-Huot, Frank Coronado, and Zahra Zahraee for useful discussions. SMC is supported by the Royal Society under the grant URF\textbackslash R1\textbackslash 221310 and the UK Engineering and Physical Sciences Research Council grant number EP/Z000106/1. The work of DRP is supported by STFC DTP research studentship grant ST/Y509231/1. AP is supported by the INFN ``Iniziativa Specifica'' ST\&FI. AP thanks the Aspen Center for Physics for its hospitality while part of this work was performed, and acknowledges its support from a grant from the Simons Foundation (1161654, Troyer). We would like to thank the organizers of Bootstrap 2026, where this work was first presented.
The computations presented here were conducted on the facilities provided by the Imperial College Research Computing Service (\url{http://doi.org/10.14469/hpc/2232}) and on the SISSA HPC cluster Ulysses.
We acknowledge the use of Claude (Anthropic) as an assistive tool for coding and for polishing the manuscript.

\appendix

\section{Mack Polynomials and Polyakov-Regge Blocks}
\label{app:mack}

\paragraph{Representation projectors.} In the planar limit the representations $\mathbf{35}_s$ and $\mathbf{35}_c$ are indistinguishable and we can thus join them in the reducible representation $\mathbf{70}$, resulting in the six representations
\es{}{
    \mathbf{28}\otimes\mathbf{28}=\mathbf{1}\oplus\mathbf{28}\oplus\mathbf{35}_c\oplus\mathbf{35}_s\oplus\mathbf{35}_v\oplus\mathbf{300}\oplus\mathbf{350}\,.
}
In the basis
\es{tbasis}{
    \mathtt{t}=\big(\delta^{ab}\delta^{cd},\,\delta^{ac}\delta^{bd},\,\delta^{ad}\delta^{bc},\,f^{ace}f^{bde},\,f^{ade}f^{bce},\,\tr(T^aT^bT^cT^d)\big)\ec
}
the projectors onto these six representations are
\begin{equation}\label{projbasis}
  \begin{aligned}
    P_{\mathbf 1}
    &=\big(\tfrac{1}{28},0,0,0,0,0\big)\cdot\mathtt{t}\ec & \qquad
      P_{\mathbf{28}}
    &=\big(0,0,0,-\tfrac{1}{12},\tfrac{1}{12},0\big)\cdot\mathtt{t}\ec\\
    P_{\mathbf{70}}
    &=\big(0,\tfrac16,\tfrac16,0,\tfrac16,-\tfrac16\big)\cdot\mathtt{t}\ec & \qquad
      P_{\mathbf{35}_v}
    &=\big(-\tfrac{1}{12},0,0,\tfrac{1}{12},-\tfrac{1}{12},\tfrac16\big)\cdot\mathtt{t}\ec\\
    P_{\mathbf{350}}
    &=\big(0,-\tfrac12,\tfrac12,\tfrac{1}{12},-\tfrac{1}{12},0\big)\cdot\mathtt{t}\ec & \qquad
      P_{\mathbf{300}}
    &=\big(\tfrac{1}{21},\tfrac13,\tfrac13,-\tfrac{1}{12},-\tfrac{1}{12},0\big)\cdot\mathtt{t}\,.
  \end{aligned}
\end{equation}
\paragraph{Mack polynomials.}
The residues entering the Polyakov-Regge block~\eqref{PBmellin} are
\es{Mackpoly}{
    \cQ^{m,\cN=2}_{\Delta,J}(u)=K^{\mathcal{N}=2}_{\Delta,J,m}\,Q_{\Delta,J,m}(u-4)\ec\qquad Q_{\Delta,J,m}(x)=-\sum_{q,k\leq J}(-m)_q\,\mu^{q,k}_{\Delta,J}\,(-x/2)_k\,,
}
with
\es{muqk}{
    \mu_{\Delta,J}^{q,k}=&\frac{J!\,\big(k+q+\tfrac{\Delta-J}{2}\big)_{J-k-q}}{k!\,(J-q+1)_q\,(J-k-q)!}\\
    &\times\sum_{p=0}^q\frac{(-1)^{k+p}\big(\tfrac{J+\Delta}{2}-p\big)_p^2\big(\tfrac12(\Delta-J-2)\big)_{q-p}\big(k-p+q+\tfrac{\Delta-J}{2}\big)_{J-k-q}}{p!\,(q-p)!\,(\Delta-J-2)_{q-p}\,(k-p+q+\Delta-1)_{J-k+p-q}}\,,
}
and \es{}{K^{\cN=2}_{\Delta,J,m}=\frac{2 \Gamma (J+\Delta -1) \Gamma (J+\Delta )}{m! \Gamma (\Delta -1) \Gamma \left(\frac{J+\Delta }{2}\right)^4 (\Delta -1)_m \Gamma \left(\frac{1}{2} (J-\Delta +6)-m\right)^2}\,.}
The double zeros on double-trace twists, which enforce the planar single-trace decoupling, originate from the coefficients $K^{\mathcal{N}=2}_{\Delta,J,m}$.
\paragraph{Large twist and flat space.} In the large-twist limit the expansion in Polyakov-Regge blocks reduces to the partial-wave expansion of the flat-space gluon amplitude in $\mathrm{AdS}$~\cite{Penedones:2010ue}. Expanding the flat-space amplitude dual to gluon scattering in partial waves, we write
\es{}{
    \mathcal{A}(s,t)=\frac{1}{s^{D/2}}\sum_J n_J\,a_J(s)\,\mathcal{P}_J\big(1+\tfrac{2t}{s}\big)\ec
}
where $D=5$ is the dimension of the flat bulk spacetime and the normalization $n_J$ and the partial waves $\mathcal{P}_J$ are given by~\cite{Correia:2020xtr}
\es{}{
    n_J=\frac{(4\pi)^{D/2}\,(D\!+\!2J\!-\!3)\,\Gamma(D\!+\!J\!-\!3)}{\pi\,\Gamma\big(\tfrac{D-2}{2}\big)\,\Gamma(J+1)}\ec\qquad \mathcal{P}_J(x)={}_2F_1\big(\!-\!J,J\!+\!D\!-\!3,\tfrac{D-2}{2},\tfrac{1-x}{2}\big)=\frac{U_J(x)}{J+1}\ed
}
The amplitude also admits a dispersive representation as a sum over string resonances of mass $m$ and spin $J$ with coefficient $C_{m,J}$, corresponding to heavy string operators,
\es{}{
    \mathcal{A}(s,t)=&\,\mathcal{A}_{\rm gluon}(s,t)+\sum_{J,m}C^2_{m,J}\big(\mathcal{P}^{\rm flat}_{m,J}(s,t)+\mathcal{P}^{\rm flat}_{m,J}(t,s)\big)\ec\\
    &\mathcal{P}^{\rm flat}_{m,J}(s,t)=\frac{n_J}{m^{D}}\,\mathcal{P}_J\Big(1-\frac{2(s+t)}{m^2}\Big)\frac{1}{m^2-s}\ec
}
and we find, as in~\cite{Caron-Huot:2024loc}, that this dispersion relation is the flat-space limit of~\eqref{OPEMellinbox}. More precisely, we find that in the large-twist limit, all orders in $s,t\ll \Delta$ of the Polyakov-Regge block~\eqref{PBmellin} in Mellin space are related to the flat-space partial waves as
\es{}{
    \frac{\lambda^{2,\text{free}}_{\Delta,J}}{2\sin^2\big(\tfrac{\pi\tau}{2}\big)}\,\cP^{\mathcal{N}=2}_{\tau,J}(s,t)\Big|_{s^at^b}=(-1)^J\,\mathcal{P}^{\rm flat}_{m,J}(s,t)\Big|_{s^at^b}\,\frac{2^{a+b-3}\Gamma(4+a+b)}{\pi^2R^{2(a+b)+5}}\,\frac{4m}{\pi R}(1+O(\Delta^{-1}))\ec
}
with $\Delta=mR$ and $R$ the $\mathrm{AdS}_5$ radius.

\section{Details of the dispersion relation}
\label{newDis}

The kernel entering the dispersion relation~\eqref{disprel} is~\cite{Carmi:2019cub,Caron-Huot:2020adz}
\es{kernelK}{
    K(U,V;U',V')=&\frac{U-V+U'-V'}{64\pi(UVU'V')^{3/4}}\,x^{3/2}\,{}_2F_1\!\big(\tfrac12,\tfrac32,2,1-x\big)\,\theta\big(\sqrt{V'}-(\sqrt U+\sqrt V+\sqrt{U'})\big)\\
    &+\frac{1}{4\pi(UVU'V')^{1/4}}\frac{1}{\sqrt U+\sqrt{U'}}\,\delta\big(\sqrt{V'}-(\sqrt U+\sqrt V+\sqrt{U'})\big)\ec
}
where
\es{xdef}{
    x=\frac{16\sqrt{UVU'V'}}{\big[(\sqrt U+\sqrt V)^2-(\sqrt{U'}+\sqrt{V'})^2\big]\big[(\sqrt U-\sqrt V)^2-(\sqrt{U'}-\sqrt{V'})^2\big]}\ed
}
The integration region corresponds to the $s$-channel cut $w,\bar w<0$, with $U'=w\bar w$ and $V'=(1-w)(1-\bar w)$.

\section{Additional details on the functionals}
\label{app:eval}
\subsection{Proof of the Mellin representation of $R$}
Starting from~\eqref{posRt}, we use that $\dDisc\, g^{\mathcal{N}=2}(u',v')=\dDisc\, \cP^{\mathcal{N}=2}(u',v')$ together with the Mellin expression~\eqref{PBmellin} for the Polyakov blocks. Doing so, we have diagonalized the double-discontinuity, which can be evaluated using \es{}{\dDisc\,u^{\tfrac{s}{2}-2}=2 \sin^2\!\big(\pi \tfrac{s}{2}\big)\,u^{\tfrac{s}{2}-2}\,.} The remaining integrals commute and the integral over $u'$ can be done analytically using
\es{}{
    \int\limits_{0}^{(\sqrt{v'}-\sqrt{v})^2}\frac{v'(u'/v')^p\,\dif{u'}}{\pi^2 u'\sqrt{\lambda(u',v',v)}}=\!\frac{\Gamma(p)\,\Gamma(p+1)}{\pi^2\,\Gamma(2p+1)}(1-t)^{2p-1} {}_2F_1(p,p,2p,1-t)\,,}
where we have set $t=v/v'$. What remains is a single hypergeometric integral over $t$, which we evaluate using the representation
\es{}{\, _3F_2\left(a_1,a_2,a_3;b_1,b_2;z\right)=\frac{\Gamma \left(b_2\right)}{\Gamma \left(a_3\right) \Gamma \left(b_2-a_3\right)} \int _0^1(1-t)^{-a_3+b_2-1} t^{a_3-1} \, _2F_1\left(a_1,a_2,b_1,t z\right)\dif{t}\,.}
Doing so, we are left with
\es{}{\int \frac{\dif{s}\,\dif{t}}{(4\pi i)^2} v^{-\tfrac{u}{2}} \Gamma(\tfrac{u}{2})^2 \Gamma(2-\tfrac{u}{2})^2\sum_m \frac{\mathcal{Q}_{\Delta +2,J}^{m,\,\cN=2}(u)}{
        (s-\Delta +J-2 m)}\,.}
The integral over $s$ can be evaluated by taking the residue at the pole, and we obtain~\eqref{Rtposmellin}, after renaming $u\to t$.

\subsection{Regge limit of the functionals}
\label{app:regge-limit}
For the stability of the numerical bootstrap it is useful to have analytic control over the
functionals at large spin and twist.
The Regge limit in the \((\tau,J)\) plane, to be defined shortly, is a particularly useful limit that we can take on the various functionals as we will now discuss.

Define the eigenvalue of the \(\cN=2\) superconformal Casimir\footnote{The case of \(\cN=4\) supersymmetry was discussed in~\cite{Caron-Huot:2024loc} and their results for the Regge limit were also used in our \(\cN=4\) bootstrap.} as
\es{}{
    m^2 = \Delta^2 - (J+1)^2
}
and consider the functionals in the limit $J \to \infty$ with
\es{}{
    \eta = 1 + \frac{2(J+1)^2}{m^2}
}
fixed. Considering the Regge limit is useful because by imposing positivity in the Regge limit for some values of \(\eta\) we are effectively imposing positivity on a large region in the \((\tau, J)\) plane. In particular, we find that imposing positivity on a grid in \(\tau\) up to \(\tau_\text{max}\) and some values of \(\eta\) produces functionals that are positive for all spins at large \(\tau\)\footnote{In fact they grow with \(\tau\) as some positive power.} even with a modest choice of \(\tau_\text{max}\).
We now report the formulas used in our setup for the functionals in the Regge limit; we find
\es{}{
    \label{ReggeRt}
    \frac{\lambda^{2\text{,\,free}}_{\Delta,J} m^2}{2\sin\left(\pi\frac{\tau}{2}\right)^2} \Rt
    = -\frac{16}{\pi^2 \eta^2} \left[
        x {}_2F_1\!\left(3,\, \tfrac{t-1}{2},\, \tfrac{3}{2},\, x\right)
        + \tfrac{1}{3} x^2 {}_2F_1\!\left(3,\, \tfrac{t+1}{2},\, \tfrac{5}{2},\, x\right)
    \right]
    + O\left(m^{-2}\right)\,,
}
and
\es{}{
    \frac{\lambda^{2\text{,\,free}}_{\Delta,J}}{2\sin\left(\pi\frac{\tau}{2}\right)^2} \Ru
    = -\frac{8\,(-1)^J}{\pi^2 \eta(u-2)}\, x {}_2F_1\!\left[2,\, \tfrac{u-1}{2},\, \tfrac{3}{2},\, x\right]
    + O\left(m^{-2}\right)\,,
}
where \(x=1-\eta^{-2}\) and $\lambda^{2,\text{free}}_{\Delta,J}$ are the generalized free field OPE coefficients,
\es{}{\lambda^{2,\text{free}}_{\Delta,J}=\frac{2 (\Delta +2) (J+1) \Gamma \left(\frac{\Delta -J}{2}+1\right)^2 \Gamma \left(\frac{J+\Delta }{2}+2\right)^2}{\Gamma (-J+\Delta +1) \Gamma (J+\Delta +3)}\,.}
An analogous formula for the \(B\) functional was proved in~\cite{Caron-Huot:2024loc} using the Regge moments; we instead make an ansatz in terms of two hypergeometric functions and fix the parameters with a numerical fit. In fixing the ansatz we exploit the analytic structure of the hypergeometric functions it contains. Expanding ${}_2F_1\big(a,\tfrac t2+b,c,x\big)$ at small $x$, the coefficient of $x^n$ is a polynomial of degree $n$ in $t$. Matching order by order in $x$ therefore determines a finite number of parameter combinations at each step and we can fix the ansatz iteratively in $n$.

From here one can find the Regge limit of the \(\Psi\) functional by integrating~\eqref{ReggeRt} against the kernel \(\Psi(t)\). The functionals \(X\) and \(I\) are sub-leading in the Regge limit with respect to the \(R\) functional and can thus be set to 0 in \texttt{SDPB}.

\subsection{Efficient evaluation of the functionals}
\paragraph{Polyakov-Regge blocks.} The infinite sum over the descendants in the definition~\eqref{PBmellin} can be evaluated exactly, using the following identity
\es{}{
    \sum_{m=0}^{\infty}K^{\mathcal{N}=2}_{\Delta+2,J,m}(-m)_q=\hat{K}^{\mathcal{N}=2}_{\Delta,J}\frac{(\tfrac{\Delta-J-2}{2})_q^2}{(-2-J)_q}\,,}
where
\es{}{
    \hat{K}^{\mathcal{N}=2}_{\Delta,J}=K^{\mathcal{N}=2}_{\Delta+2,J,0}\frac{\Gamma(J+3)\Gamma(\Delta+1)}{\Gamma^2(\tfrac{\Delta+J+4}{2})}\,.
}
Following a similar strategy to~\cite{Caron-Huot:2022sdy}, we find that we can write
\es{}{\label{efficientPolyakov}
    \mathcal{P}^{\mathcal{N}=2}_{\tau,J}(s,t)=(-1)^{J+1}\hat{K}^{\mathcal{N}=2}_{\Delta,J}\sum_{q,k}\hat{P}^{\mathcal{N}=2}_{\Delta,J,q}(s)\mu_{\Delta+2,J}^{q,k}(\tfrac{s+t-2}{2})_k\,,
}
where the coefficients $\hat{P}^{\mathcal{N}=2}_{\Delta,J,q}(s)$ are given by the recursion relation\footnote{
    We found this recursion relation, as well as the \( \cN = 4 \) analogue, to be numerically unstable at large \( J \). Up to a rescaling it can be cast into a recursion of the schematic form \( p_{q,J} = p_{q-1,J} + c_{q,J}(s,\tau) \) whose solution is \( p_{q,J} = p_{0,J} + \sum_{k=1}^{q} c_{k,J}(s,\tau) \). It turns out that for large \( J \) and \( q \sim J/2 \), the sum is very ill-conditioned as \( p_{0,J} \approx - \sum_{k=1}^{q} c_{q,J}(s,\tau) \). It is then necessary to evaluate this recursion relation at high numerical precision and benchmark the digit loss to obtain a reliable value.
}
\es{}{
    \hat{P}^{\mathcal{N}=2}_{\Delta,J,q}(s)=\frac{\Delta-J-s+2q-2}{2}\hat{P}^{\mathcal{N}=2}_{\Delta,J,q-1}(s)+\frac{(\tfrac{\Delta-J-2}{2})_{q-1}}{2(-J-2)_{q-1}}
    \,,}
with the seed of the recursion being
\es{}{
    \hat{P}^{\mathcal{N}=2}_{\Delta,J,0}(s)=\frac{\Gamma^2(\tfrac{\Delta+J+4}{2})}{\Gamma(J+3)\Gamma(\Delta+1)}\frac{\, _3F_2\left(\tfrac{\Delta-J-2}{2},\tfrac{\Delta-J-2}{2},\tfrac{\Delta-J-s}{2};\tfrac{\Delta-J-s+2}{2},\Delta +1;1\right)}{s-\Delta+J}\,.
}
The evaluation of the seeds of the recursion is numerically very expensive using the standard Mathematica implementation. It can be made much faster by noticing that
\es{}{
    \frac{(2 J\!+\!\tau \!+\!4) \Gamma(\tfrac{2J+\tau+4 }{2})^2 {} _3F_2(\tfrac{\tau-2
        }{2},\frac{\tau -2}{2},\frac{\tau -s}{2};\frac{\tau-s+2}{2},J\!+\!\tau \!+\!1;1)}{2 \Gamma (J+4) \Gamma (J+\tau +1)}= {}_3F_2(a,b,1;a+n,a+b;1)\,,
}
where
\begin{equation}
    a=\frac{\tau-2}{2}\ec \qquad b=2-\frac{s}{2} \ec \qquad n=J+4 \ed
\end{equation}
The latter \( {} _3F_2 \) can be efficiently evaluated using a three-term recursion relation~\cite[\href{https://dlmf.nist.gov/16.4.E12}{Eq.\ 16.4.12}]{NIST:DLMF}
\es{}{{}_3F_2(1,a,b;a+b,a+n;1) &= \tfrac{(a+n-1)  (a^2+3 a (n-2)-(n-2) (b-2 n+4))}{(n-1) (a+n-2) (a-b+n-1)}{}_3F_2(1,a,b;a\!+\!b,a\!+\!n\!-\!1;1)\\
    & \quad  + \tfrac{(3-a-n) (a+n-1) }{(n-1) (a-b+n-1)}{}_3F_2(1,a,b;a\!+\!b,a\!+\!n\!-\!2;1)\,.}
To evaluate this, we need the \( n = 0 \) term, which is trivial, and the \( n = 1 \) term, which we compute with Mathematica's \texttt{HypergeometricPFQ} and remains the only expensive part of the computation of the \( \hat{P}^{\mathcal{N}=2}_{\Delta,J,q} \).

To complete the computation of the Polyakov block, we need the coefficients \( \mu_{\Delta+2,J}^{q,k} \). Computing them from~\eqref{muqk} is quite inefficient, and a much better strategy is to use again a recursion relation, which is discussed in~\cite{Caron-Huot:2022sdy}. Finally, the sum~\eqref{efficientPolyakov} is also badly ill-conditioned because of alternating signs in the Pochhammer symbols and the \( \mu_{\Delta+2,J}^{q,k} \) that lead to large cancellations. At least in this case, we can circumvent this problem by evaluating \( \mu_{\Delta+2,J}^{q,k} \) and \( (\tfrac{s+t-2}{2})_k \) at rational values and performing the sum over \( k \) analytically. The remaining sum over \( q \) is then well-conditioned.

\paragraph{Localization functional.} We now describe how to compute the localization integral~\eqref{IntKernel2} applied to the Polyakov-Regge blocks using the finite sum expression~\eqref{efficientPolyakov}.

Our goal will be to evaluate the integral over $t$ analytically so that we are left with a single Mellin-Barnes integral in $s$ that can be evaluated numerically to high precision thanks to the decaying behavior of the gamma functions. We see that we have 5 different integrals to evaluate
\es{}{
    \int \frac{\dif{t}}{2\pi i} \Gamma(2-\tfrac{t}{2}) \Gamma(\tfrac{t}{2}) \Gamma(\tfrac{6-s-t}{2}) \Gamma(\tfrac{s+t-2}{2} +k) f_i(s,t)\,,
}
where
\es{}{f_i=\left\{\frac{1}{(t\!-\!2) (s\!+\!t\!-\!4)},\frac{H_{1-\tfrac{t}{2}}}{(s\!-\!2) (s\!+\!t\!-\!4)},\frac{H_{\tfrac{4-s-t}{2}}}{(t\!-\!2) (s\!-\!2)},\frac{H_{\tfrac{t}{2}-1}}{(s\!-\!2) (s\!+\!t\!-\!4)},\frac{H_{\tfrac{s+t-4}{2}}}{(t\!-\!2) (s\!-\!2)}\right\}\,.}
The first integral can be done with the first Barnes lemma
\es{}{
    \int \frac{\dif{s}}{2 \pi i}\Gamma(a+s)\Gamma(b+s)\Gamma(c-s)\Gamma(d-s)=\frac{\Gamma(a+c)\Gamma(a+d)\Gamma(b+c)\Gamma(b+d)}{\Gamma(a+b+c+d)}\,.
}
The integrals involving the harmonic number can be evaluated as follows. Observe that
\es{}{H_s=\psi^{(0)}(s+1)+\gamma\qquad \text{and}\qquad \left.\frac{\mathrm{d}}{\mathrm{d}\epsilon}\frac{\Gamma(s+1+\epsilon)}{\Gamma(s+1)}\right\rvert_{\epsilon=0}=\psi^{(0)}(s+1)\,,}
we can therefore focus on the integral
\es{}{
    \int \frac{\dif{s}}{2 \pi i}\Gamma (a+s) \Gamma (b+s) \Gamma (c-s) \Gamma (d-s) \frac{\Gamma (\epsilon+s+1)}{\Gamma (s+1)}
    \,,}
and use the representation \es{}{\int_0^1 \dif{x} \,\frac{x^{-1-\epsilon}(1-x)^{s+\epsilon}}{\Gamma(-\epsilon)}=\frac{\Gamma (\epsilon+s+1)}{\Gamma (s+1)}\,.}
We are left with a simple Barnes integral in $s$ that can be evaluated as
\es{}{
    (1-x)^{c+\epsilon}x^{-1-\epsilon}\Gamma(a+c)\Gamma(b+c)\Gamma(a+d)\Gamma(b+d)\frac{{}_2F_1(a+c,b+c,a+b+c+d,x)}{\Gamma(a+b+c+d)\Gamma(-\epsilon)}\,,
}
and the integral over $x$ can now be evaluated using a representation of ${}_3F_2$. Taking the derivative with respect to $\epsilon$ and setting $\epsilon$ to 0, we find
\es{}{
    &\int_{-i \infty}^{i \infty}\frac{\dif{s}}{2 \pi i}\Gamma(a+s)\Gamma(b+s)\Gamma(c-s)
    \Gamma(d-s)H_s \\
    & = \frac{\Gamma(a\! + \!c) \Gamma(b\! + \!c) \Gamma(a\! +\! d) \Gamma(b\! +\! d) }{\Gamma(a+b+c+d)}\left[H_c\!-\!\left.\frac{\mathrm{d}}{\mathrm{d}\epsilon} {}_3F_2(a\!+\!c,b\!+\!c,\epsilon;1\!+\!c,a\!+\!b\!+\!c\!+\!d;1)\right\rvert_{\epsilon=0}\right]
}
The remaining term is evaluated by differentiating the defining series of ${}_3F_2$ term by term. The resulting sum can be resummed in closed form, giving
\es{}{
    &\left.\frac{\mathrm{d}}{\mathrm{d}\epsilon}{}_3F_2(a\!+\!c,b\!+\!c,\epsilon ;1\!+\!c,a\!+\!b\!+\!c\!+\!d;1)\right\rvert_{\epsilon=0} \\
    &\qquad = \frac{(a+c)(b+c)}{(1\!+\!c)(a\!+\!b\!+\!c\!+\!d)}{}_4F_3(1,1,1\!+\!a\!+\!c,1\!+\!b\!+\!c;2,2\!+\!c,1\!+\!a\!+\!b\!+\!c\!+\!d;1)\,.
}
Having proved this lemma, it is easy to show that
\es{integratedefficient}{
    \Itwo=\hat{K}^{\mathcal{N}=2}_{\Delta,J}(-1)^{J+1} \!\int\!\frac{\dif{s}}{16 \pi i}\Gamma^2(\tfrac{s}{2})\Gamma^2(1-\tfrac{s}{2})\sum_{q,k}\hat{P}^{\mathcal{N}=2}_{\Delta,J,q}(s)\mu_{\Delta+2,J}^{q,k}I^{\mathcal{N}=2}_k(s)
    \,,}
where
\es{}{
    I^{\mathcal{N}=2}_k(s)=
    \frac{1}{(k+1)(k+2)}\Big[(k+1)!+\big(\tfrac{s}{2}-1\big)_{k+2}\big(2H_{k+1}-H_{k+\tfrac{s}{2}}+H_{\tfrac{s}{2}-2}\big)\Big]\,.}
The integral in~\eqref{integratedefficient} is computed numerically as an exponentially convergent Riemann sum after the change of variable \( s = 1 + i \sinh(x) \), as done in the \( \cN = 4 \) case in~\cite{Caron-Huot:2024loc}. The terms \( \hat{P}^{\mathcal{N}=2}_{\Delta,J,q}(s) \) and \( \mu_{\Delta+2,J}^{q,k} \) are computed in the same way as for the Polyakov blocks. In this case, we cannot avoid the ill-conditioning in the sum over \( k \) since we are not able to keep \(  I^{\mathcal{N}=2}_k(s) \) rational. We are thus forced to perform this computation at very high numerical precision.

\paragraph{Regge functionals.} The infinite sum over $m$ in the definition of the Regge functionals can also be resummed exactly, giving
\es{}{
    \Ru &=-(-1)^J\,\hat K^{\mathcal{N}=2}_{\Delta,J}\sum_{q,k}\hat R^u[\tau,J,u,q]\,\mu^{q,k}_{\Delta+2,J}\big(\tfrac{4-u}{2}\big)_k\ec \\
    \Rt &=-\hat K^{\mathcal{N}=2}_{\Delta,J}\sum_{q,k}\hat R^t[\tau,J,q]\,\mu^{q,k}_{\Delta+2,J}\big(\tfrac{4-t}{2}\big)_k\,,
}
where
\es{}{
    \hat{R}^u[\tau,J,u,q]&=\frac{1}{u-2}\frac{2 \left(\frac{\tau -2}{2}\right)_q^2 \left(2 q+\tau+\frac{u-4}{2}-1\right)}{(-J-2)_q}-\frac{4 \left(\frac{\tau-2}{2}\right)_{q+1}^2}{(-J-2)_{q+1}}\,, \\
    \hat{R}^t[\tau,J,q]&=\frac{(-1)^q \left(\frac{1}{2} (-2 q-\tau +4)\right)_q^2}{(J-q+3)_q}
    \,.
}

Integrating the kernel of the projection functional $\Psi$ against the Pochhammer symbols gives
\es{}{ \int \frac{\dif{t}}{2 \pi i} \Psi_\ell(t) \big(\tfrac{4-t}{2})_k =-\frac{(-1)^\ell 2^{-2 (\ell+1)} (\ell+1) \Gamma (k+1) \Gamma (k+2) \Gamma (\ell+3)}{\left(\frac{3}{2}\right)_\ell (k+2)_{\ell+2} \Gamma (k-\ell+1)}\equiv [\Psi_\ell]_k\,,}
allowing us to easily compute the functionals on the blocks.

\subsection{Large spin limit of the functionals}
The cost of evaluating the functionals grows as $J^2$, which in practice limits the maximal spin we can include in the numerics. We therefore supplement the explicit data with the large-spin behavior of the functionals at fixed twist, which we now discuss. The analogous asymptotics for the $\mathcal{N}=4$ functionals were obtained in~\cite{Caron-Huot:2022sdy}, where they were used to establish the domain of convergence of the corresponding sum rules~\cite{Alday:2022uxp}.

Let us normalize the functionals with
\es{}{\cN_{\tau,J}=\frac{1}{8} \pi ^2 \frac{(\tau -2)^2}{\sin(\tau\pi/2)^2} (\tau +1)^2 (2 J+\tau +3)^2 \lambda^{2,\text{free}}_{\Delta,J}\,.}
For the Polyakov-Regge blocks we find that
\begin{equation}
    \begin{aligned}
      \cN_{\tau,J} \cP^{\mathcal{N}=2}_{\tau,J}(s,t)&\sim J^{7 - s - t} \quad && \text{for} \ 0<s,t<2\,, \\
      \cN_{\tau,J} \cP^{\mathcal{N}=2}_{\tau,J}(t,6-s-t)&\sim J^{7-s - \tau}+J^{s + 1}  \quad && \text{for} \  0<s,t<2\,.
    \end{aligned}
\end{equation}
For the $R^t$ functionals, we find
\es{}{
    \cN_{\tau,J} \Rt \sim J^{7 - \tau - t}+J^{1 + t}  \quad \text{for\,} 0<t<2 \,.
}
Lastly, for the integrated functional $\Itwo$, we are able to fix the first few coefficients in the large-$J$ expansion
\es{}{
    \cN_{\tau,J}\Itwo =(-1)^J\frac{(\tau +1)^2}{\tau } \left(J \log (J)-J \psi\left(\frac{\tau }{2}\right)+\left(\frac{\tau }{2}+5\right) \log (J)+O(1)\right)\,.
}
In the $\cN=2$ bootstrap setup we can therefore compute the functionals up to a large value of $J$, say $J=300$, and then extrapolate with a simple fit in $1/J$ to obtain a satisfactory approximation of the functionals for even higher spin, thus reducing significantly the computational resources used to compute the functionals.

\begin{figure}[t]
    \centering
    \begin{tikzpicture}[line width=0.9pt, scale=1.05]
        \fill[gray!15] (0,2) -- (2,0) -- (4,0) -- (4,2) -- (2,4) -- (0,4) -- cycle;
        \begin{scope}[gray!60, dashed, line width=0.6pt]
            \foreach \x in {2,4}{ \draw (\x,-0.75) -- (\x,5.1); }
            \foreach \y in {2,4}{ \draw (-1.15,\y) -- (5.35,\y); }
            \draw (-0.55,4.55) -- (4.55,-0.55);
        \end{scope}
        \draw (0,2) -- (2,0) -- (4,0) -- (4,2) -- (2,4) -- (0,4) -- cycle;
        \draw[->] (-1.4,0) -- (5.9,0) node[below left] {$\Re(s)$};
        \draw[->] (0,-1.0) -- (0,5.6) node[below right] {$\Re(t)$};
        \foreach \x in {2,4}{
            \draw (\x,0.12) -- (\x,-0.12);
            \node[fill=white, inner sep=1.5pt] at (\x,-0.42) {$\x$};
        }
        \foreach \y in {2,4}{
            \draw (0.12,\y) -- (-0.12,\y);
            \node[fill=white, inner sep=1.5pt] at (-0.42,\y) {$\y$};
        }
        \node at (2.7,2.7) {$\cC$};
        \pgfmathsetmacro{\redmarg}{0.08}
        \pgfmathsetmacro{\reddiag}{\redmarg*(1+sqrt(2))}
        \begin{scope}[red, line width=0.7pt]
            \draw ({\reddiag},{2-\redmarg}) -- ({2-\redmarg},{2-\redmarg})
                  -- ({2-\redmarg},{\reddiag}) -- cycle;
            \draw ({2+\reddiag},{2-\redmarg}) -- ({4-\redmarg},{2-\redmarg})
                  -- ({4-\redmarg},{\reddiag}) -- cycle;
            \draw ({\reddiag},{4-\redmarg}) -- ({2-\redmarg},{4-\redmarg})
                  -- ({2-\redmarg},{2+\reddiag}) -- cycle;
            \node at (1.35,1.35) {$\cC_u$};
            \node at (3.35,1.35) {$\cC_s$};
            \node at (1.35,3.35) {$\cC_t$};
        \end{scope}
    \end{tikzpicture}
    \caption{The region $\cC$ in the plane of the real parts of the Mellin variables.}
    \label{fig:convergence}
\end{figure}
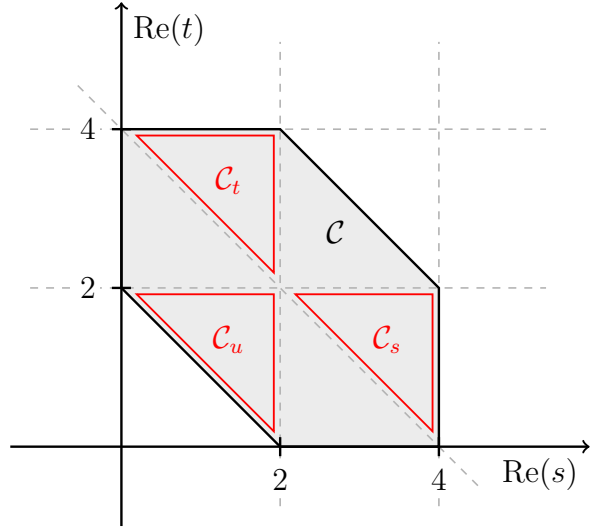

\section{Region of convergence}\label{app:convergence}
The Mellin representation of the correlator~\eqref{MellinN2} is valid in the region $\cC$ that separates the poles of the Gamma functions, represented in Figure~\ref{fig:convergence}.

Poles in the Mellin amplitude can be present in the region \(\cC\), and they can in principle introduce discontinuities as residues are picked up when moving in the region. This is the case for the supergravity amplitude and the 1-loop amplitude that have poles at \(s=2\) and \(t=2\), and for the Polyakov block associated with operators with twist \(2<\tau<4\).
The solution in our case is straightforward as the region \(\cC_u=\cC\cap \{\Re(s)<2,\Re(t)<2\}\) is guaranteed to be free of singularities, and its images under crossing cover the entire region \(\cC\).

Working in \(\cC_u\) leads us naturally to the domain of convergence of the Polyakov blocks expansion~\eqref{OPEMellinbox}, the \( X \), the \(R\), and \(\tilde{R}\) sum rules in~\eqref{crossmaster},~\eqref{Rtsum}, and~\eqref{Rusum}, respectively. Similar considerations on the $\cN=4$ correlator lead to the domain of convergence of the Polyakov expansion~\eqref{OPEMellinbox4} and the sum rules~\eqref{crossmaster4} and~\eqref{ReggeN4}~\cite{Caron-Huot:2020adz}. This motivates the choice of sampling points in equations~\eqref{stListN4},~\eqref{stListN2},~\eqref{tListN4}, and~\eqref{tListN2}. Note, in particular, that for~\eqref{crossmaster4} we sample only ``half'' of the triangle \( \{(s,t) : \Re(s),\Re(t),\Re(\tilde{u}) < 6\} \) since \( \Xfour = X^{\cN=4}_{16-s-t,t}[\tau,J] \).

\section{Checks on the $\mathcal{N}=2$ integrated constraint}
\label{weak}

The evaluation of the integral~\eqref{IntKernel2} requires some care if the Mellin amplitude has poles, as is the case for the weak and strong coupling limits of the amplitude~\eqref{weakN2}, \eqref{strongN2}. Let us define the region
\es{}{
    \cC=\{(s,t) \colon 0<\Re(s)<4, \ 0<\Re(t)<4, \ 0<\Re(u)<4\}\,,
}
where the kernel of the integral has no discontinuities. The integral must be evaluated on vertical contours in the complex $(s,t)$ variables chosen so that the $s$- and $t$-poles of the amplitude lie to the right of the contour and the $u$-poles to the left. We cannot find values of ${\rm Re}(s)$ and ${\rm Re}(t)$ that satisfy this prescription for $\nobreak{M(s,t)+M(t,u)+M(u,s)}$ if $M(s,t)$ has poles at $s=2$ and $t=2$, as is the case for the weak and strong coupling amplitudes, so we split the integral and integrate on the following domains,
\es{}{
    &\mathcal{I}[M(s,t)]|_{s,t\in \cC_u}\,,\\
    &\mathcal{I}[M(t,u)]|_{s,t\in \cC_s}\,,\\
    &\mathcal{I}[M(u,s)]|_{s,t\in \cC_t}\,,
}
where the contours are chosen as
\es{}{
    \cC_u&=\cC\cap \{\Re(s)<2,\Re(t)<2\},\\
    \cC_s&=\cC\cap \{\Re(t)<2,\Re(u)<2\},\\
    \cC_t&=\cC\cap \{\Re(u)<2,\Re(s)<2\}\,.
}
Evaluating both sides of~\eqref{locN2} at order $\lambda^1$, we find
\es{}{
    \frac{4\pi}{\sqrt{\lambda}}\int_{0}^{\infty} \frac{\omega^{2}\,\dif{\omega}}{(\sinh \omega)^{2}}\,\left. J_1(\tfrac{\sqrt{\lambda}\omega}{\pi})\right\rvert_{\lambda^1}&=-\frac{15}{8} \frac{\zeta_{5}}{\pi^2}\lambda=
    \mathcal{I}[M(s,t)+M(t,u)+M(u,s)]\Big\rvert_{\lambda^1}
    \,.
}
The situation is similar at strong coupling. Naively, one would find $M(s,t)+M(t,u)+M(u,s)=0$ and conclude that the integral vanishes, but this manipulation is not allowed: the poles of $M(s,t)$ obstruct the choice of a common contour for the three terms, so there is no domain on which the cancellation can be carried out under the integral sign. It would moreover be incompatible with the position-space result~\eqref{Tgluonfull}, which has a non-vanishing $\mathtt{t}_6$ component.

Splitting the integral as described, we find $\mathcal{I}[M(s,t)]\!=\!\mathcal{I}[M(t,u)]\!=\!\mathcal{I}[M(u,s)]\!=\!-\zeta_3$, showing that the contribution from $M_\text{gluon}$ cancels the free-theory part of the localization input to reproduce~\eqref{opeconstraint}.

\bibliographystyle{JHEP}
\bibliography{planarBoot.bib}

\end{document}